\documentclass[twocolumn,           
               showpacs,            
               preprintnumbers,     
               aps,                 
               prd,                 
               superscriptaddress,      
               nofootinbib,         
               tightenlines,        
               floats,floatfix      
               ]{revtex4-1}

\usepackage{graphicx}
\usepackage{color}
\usepackage{latexsym}
\usepackage{amsmath,amssymb}        
\usepackage[draft=false]{hyperref}

\begin{document}


\title{Revisiting spherically symmetric relativistic hydrodynamics}

\author{F. S. Guzm\'an, F. D. Lora-Clavijo, M. D. Morales}
\affiliation{Instituto de F\'{\i}sica y Matem\'{a}ticas,
        Universidad Michoacana de San Nicol\'as de Hidalgo. Edificio C-3,
        Cd. Universitaria,
        C. P. 58040 Morelia, Michoac\'{a}n, M\'exico.}


\date{\today}


\begin{abstract}
In this paper we revise two classical examples of Relativistic Hydrodynamics in order to illustrate in detail the numerical methods commonly used in fluid dynamics, specifically those designed to deal with shocks, which are based on a finite volume approximation. The two cases we consider are the relativistic blast wave problem and the evolution of a Tolman-Oppenheimer-Volkoff star model, in spherical symmetry. In the first case we illustrate the implementation of relativistic Euler's equations on a fixed background space-time, whereas in the second case we also show how to couple the evolution of the fluid to the evolution of the space-time.
\end{abstract}




\maketitle



\section{Introduction}

In this paper we present the elements needed to implement the numerical solution to the relativistic Euler equations in spherical symmetry using examples in special and general relativity. The aim is to describe the necessary tools to implement numerical codes able to deal with basic problems involving relativistic hydrodynamics which eventually are used to model high energy astrophysical phenomena,  for instance stellar collapse of compact stars like supernovae core collapse, shock waves propagating out from a spherical source interacting with the interstellar medium, etc. We are particularly interested in revising  the specific numerical methods that are commonly used and present a detailed flavor of high resolution shock capturing methods used in general relativistic hydrodynamics. We focus on the description of two representative physical cases: the spherically symmetric blast wave on Minkowski space-time, which is considered to be a test case in hydrodynamics and the evolution of a Tolman-Oppenheimer-Volkoff (TOV) star model, made of a self-gravitating polytropic ideal gas on a dynamical space-time background. 

We consider the problem of solving relativistic hydrodynamical systems as an initial value problem, ruled by relativistic Euler's equations. We have to provide initial data for a relativistic fluid, which then evolves  according to Euler's equations in the blast wave case, and in the case of the TOV star we need to solve simultaneously Euler's and Einstein's equations.

In the blast wave case  we start up with free initial data, which we choose to correspond to an ideal gas  distributed into two concentric spherical chambers, being the inner one where the gas is at high pressure whereas in the outer sphere the pressure is smaller. In the TOV star case it is not possible to choose arbitrary initial data, it is necessary to construct initial data that are consistent with Einstein's equations. In order to have a complete description of the second problem we introduce the necessary general relativistic background.

We base our numerical treatment on an Eulerian description of the fluid equations using a flux balance form of the system of equations, which requires the definition of conservative variables on top of the discrete cell centered mesh. The solution of relativistic Euler's equations is found using a High Resolution Shock Capturing method (HRSC) based on the approximate solution of local Riemann problems at the intercell boundaries. Particularly we use the HLLE Riemann approximate solver with a linear piecewise reconstructor of variables. 
Being this a paper revisiting the numerical methods, we try to be as specific as possible in the description of each of the steps within the appropriate section.

The paper is organized as follows. In section \ref{sec:hydro} we set the equations of hydrodynamics for a  spherically symmetric space-time, define conservative variables and set the system of Euler's equations as a flux balance set of equations, so as the numerical methods used for the solution. In \ref{sec:blast} we present the blast wave case and describe in detail its properties and in \ref{sec:tov} we present  the evolution of TOV stars. Finally in \ref{sec:comments} we present some final comments.


\section{Hydrodynamics}
\label{sec:hydro}

\subsection{The equations of relativistic hydrodynamics in spherical symmetry}

As a starting point we describe a spherically symmetric space-time line element in spherical coordinates to be of the form

\begin{equation}
ds^2 = -\alpha^2(t,r) dt^2 +a^2(t,r)dr^2 + r^2 d\theta^2 + r^2 \sin^2 \theta d\phi^2, \label{eq:metric}
\end{equation}

\noindent where $t$ is the time coordinate and $(r,\theta,\phi)$ are the usual spherical coordinates and where we have assumed geometric units where $G=c=1$. This line-element will serve to workout the two problems we deal with: the hydrodynamics onto the Minkowski space-time where $\alpha=a=1$ and the TOV star where such metric functions obey Einstein's equations. We choose the matter model to correspond to a perfect fluid, which means the fluid is not subject to heat transfer and viscosity effects; the stress energy tensor of a perfect fluid given in general relativistic form reads \cite{Baumgarte,Schutz}:

\begin{equation}
T^{\mu\nu} = \rho_0 h u^{\mu}u^{\nu} + p g^{\mu\nu},
\label{eq:set_pf}
\end{equation}

\noindent where $\rho_0$ is the rest mass density of the fluid,  $p$ its pressure,  $u^{\mu}$ is the four velocity of the fluid on the space-time described by (\ref{eq:metric}) and $h$ the specific enthalpy

\begin{equation}
h=1+\epsilon + p/\rho_0,\label{eq:enthalpy}
\end{equation}

\noindent where $\epsilon$ is the specific internal energy of the gas. The stress-energy tensor (\ref{eq:set_pf}) is commonly found in the literature written as $T^{\mu\nu} = (\rho+p)u^{\mu}u^{\nu} + p g^{\mu\nu}$, where $\rho$ is the total energy density of the gas which can be expressed in terms of the rest mass density and internal energy as $\rho=\rho_0 (1+\epsilon)$.  Also $g^{\mu\nu}$ are the components of the inverse of the 4-metric in (\ref{eq:metric}).

The fluid equations are given by the local mass conservation law and the Bianchi identity, that are respectively:

\begin{eqnarray}
\nabla_{\mu}(\rho_0 u^{\mu}) &=& 0, \label{eq:conservation}\\
\nabla_{\mu }T^{\mu\nu} &=& 0, \label{eq:Bianchi}
\end{eqnarray}

\noindent where $\nabla_{\mu}$ is the covariant derivative consistent with (\ref{eq:metric}). When these equations are projected onto space-like hypersurfaces and their normal directions one obtains the relativistic Euler equations \cite{valencia1997,FontMiller,Alcubierre}. The result is a set of equations for the primitive variables $\rho_0,v^r,p$ or equivalently $\rho_0,v^r,\epsilon$, where $v^r$ is the three velocity of the fluid elements measured by an Eulerean observer. The way to relate the spatial velocity with the spatial components of the four velocity of the fluid in (\ref{eq:set_pf}) is using the relation $v^r = u^r \sqrt{1-g_{rr} v^r v^r}=u^r \sqrt{1-a^2 v^r v^r}=u^r/W$, where $W$ is the Lorentz factor, which in turn is defined by $W=\alpha u^t$.

It is well known that Euler's equations develop discontinuities in the hydrodynamical variables even if smooth initial data are considered \cite{LeVeque}. Therefore one may use as a first try a finite differences approach, nevertheless it cannot be applied because the approximations of derivatives would not be accurate at discontinuities; even though it is common to use finite differences modifying Euler's equations with a dissipative term and analyze the limit at which such term vanishes \cite{LeVeque}. Instead, a more accurate approach used to solve hydrodynamics equations considers the use of finite volume methods, which need the system of equations to be written in a flux balance law form, which in turn requires the definition of conservative variables as shown below.

As an illustration of how to write down a balance flux equation we construct the first of Euler's equations, the one obtained by developing (\ref{eq:conservation}) for the line element (\ref{eq:metric}):

\begin{eqnarray}
\nabla_{\mu}(\rho_0 u^{\mu}) = \frac{1}{\sqrt{-g}} \partial_{\mu} (\sqrt{-g}\rho_0 u^{\mu}) &=& 0, ~~~ \Rightarrow\nonumber\\
\frac{1}{\alpha a r^2 } \partial_t (\alpha a r^2 u^t) + \frac{1}{\alpha a r^2} \partial_r (\alpha a r^2 u^r) &=& 0.
\end{eqnarray}

\noindent where $g$ is the determinant of the metric tensor in (\ref{eq:metric}) and therefore $\sqrt{-g}=\alpha a r^2 \sin \theta$. Defining $D=\rho_0 W = \rho_0 \alpha u^t$ and considering $\alpha a \ne 0$, we have

\begin{eqnarray}
\partial_t (\rho_0 \alpha a u^t) + \frac{1}{r^2}\partial_r (\rho_0 \alpha a r^2 u^r) &=&0,~~~ \Rightarrow \nonumber\\
\partial_t (\rho_0 \alpha a u^t) + \frac{1}{r^2}\partial_r (\rho_0 \alpha a r^2 [\alpha v^r u^t]) &=&0, ~~~ \Rightarrow \nonumber\\
\partial_t (aD) + \frac{1}{r^2}\partial_r (\alpha a r^2 v^r D)&=&0,
\end{eqnarray}

\noindent which is the first of Euler's equations. The remaining equations are obtained in a similar way by developing (\ref{eq:Bianchi}), which is shown \cite{MapleLeaf_eqs}. Finally one obtains the following set of equations

\begin{eqnarray}
\partial_t (aD) &+& \frac{1}{r^2} \partial_r (\alpha a r^2 D v^r) = 0,\nonumber\\
\partial_t (aS_r) &+& \frac{1}{r^2} \partial_r (\alpha a r^2 [S_r v^r + p]) = 
\alpha a \frac{2p}{r}\nonumber\\
&-&\alpha a \frac{a^2 m}{r^2}(S_r v^r + \tau + p + D), \nonumber\\
\nonumber\\
\partial_t (a \tau) &+& \frac{1}{r^2}\partial_r (\alpha a r^2 (\tau + p)v^r) = -\alpha a \frac{m}{r^2} S_r,\label{eq:evolution_euler}
\end{eqnarray}

\noindent where the set of conservative variables is defined by

\begin{eqnarray}
D &=& \rho_0 W,\nonumber\\
S_r &=& \rho_0 h W^2 v_r, \nonumber\\
\tau &=& \rho_0 h W^2 - p - \rho_0 W. \label{eq:cons_vars}
\end{eqnarray}

\noindent Then it is possible to write down these equations as a set of balance flux type of equations

\begin{equation}
\partial_t (a{\bf u}) + \frac{1}{r^2}\partial_r (\alpha a r^2 {\bf F(u)}) = {\bf S(u)}, \label{eq:FluxConservative}
\end{equation}

\noindent where ${\bf u}$ is the state vector of conservative variables, ${\bf F}$ the flux vector and ${\bf S}$ is a source vector given by

\begin{eqnarray}
{\bf u} &=& 
	\left[
	\begin{array}{c}
	D\\ 
	S_r \\ 
	\tau 
	\end{array}
	\right], ~~~
{\bf F}. = 
	\left[
	\begin{array}{c}
	Dv^r\\
	S_r v^r +p \\
	(\tau + p)v^r
	\end{array}
	\right], \nonumber\\
{\bf S} &=&
	\left[
	\begin{array}{c}
	0\\
	-\alpha a \frac{a^2 m}{r^2}(S_r v^r + \tau + p + D) + \alpha a \frac{2p}{r}\\
	-\alpha a \frac{m}{r^2} S_r
	\end{array}
	\right].
\end{eqnarray}

Notice that the term that goes as $\sim p/r$ is singular at $r=0$. However in order to regularize the equations there, it is possible to split the flux balance form of the equations by appropriately splitting the flux vector and avoid the presence of such singular term \cite{HawkeMillmore,NielsenChoptuik2000}:

\begin{equation}
\partial_t {(a\bf u}) + \frac{1}{r^2}\partial_r (\alpha a r^2 {\bf f_1(u)}) + \partial_r (\alpha a {\bf f_2 (u)}) = {\bf s(u)}, \label{eq:FluxConservative2}
\end{equation}

\noindent where now the fluxes read:

\begin{eqnarray}
{\bf f_1} &=& 
	\left[
	\begin{array}{c}
	Dv^r\\s
	S_r v^r\\
	(\tau + p)v^r
	\end{array}
	\right], ~~~
{\bf f_2} = 
	\left[
	\begin{array}{c}
	0\\
	p \\
	0
	\end{array}
	\right], \nonumber\\
{\bf s} &=&
	\left[
	\begin{array}{c}
	0\\
	-\alpha a \frac{a^2 m}{r^2}(S_r v^r + \tau + p + D)\\
	-\alpha a \frac{m}{r^2} S_r
	\end{array}
	\right].\label{eq:conservative_split_flux}
\end{eqnarray}

\noindent There is still a usual ingredient when solving problems in spherical symmetry, that is, the coordinate singularity at $r=0$ of the derivative operators  in (\ref{eq:FluxConservative2}). This problem is solved usually by substituting $\frac{1}{r^2}\partial_r f$  by $3\partial_{r^3}f$ for a given function $f$,  where now the derivative is with respect to $r^3$. This result is applied to the second term in equation (\ref{eq:FluxConservative2}).

Therefore, we end up with three evolution equations for the four variables $D,S_r,\tau,p$, or equivalently, for the primitive variables $\rho_0,v^r,\epsilon,p$. This requires to close the system, for which an equation of state relating $p=p(\rho)$ is sufficient. In the problems we deal with in this paper we choose the gas to obey an ideal gas equation of state given by 

\begin{equation}
p= (\Gamma -1) \rho_0 \epsilon,     \label{eq:eos}
\end{equation}

\noindent where $\Gamma=c_p / c_v$ is the ratio between the specific heats, sometimes described in terms of the polytropic index $n$ such that $\Gamma =1 + 1/n$ \cite{Shapiro}. 


\subsection{Spectral decomposition of the spherically symmetric relativistic Euler equations}

The High Resolution Shock Capturing methods used here consider schemes where the spectral elements of the Jacobian matrix, ${\bf A}({\bf u})=\frac{\partial{\bf F}({\bf u})}{\partial{\bf u}}$, associated to relativistic Euler's equations (\ref{eq:FluxConservative}) play an important role. Following \cite{FontMiller}, the three eigenvalues of the Jacobian matrix of Euler's equations are as follows:

\begin{eqnarray}
&&\lambda_1 = - \beta^r + \alpha v^r  \label{eq:eigenvalues}\\
&&\lambda_{2} = -\beta^r + \frac{\alpha}{1-v^2 c_{s}^{2}} \times\nonumber\\
&&	\left[v^r (1- c_{s}^{2}) 
	+ \sqrt{c_{s}^{2} (1-v^2)[g^{rr}(1-v^2c_{s}^{2})
							-v^r v^r (1-c_{s}^{2})]
							} \right] \nonumber \\
&&\lambda_{3} = -\beta^r + \frac{\alpha}{1-v^2 c_{s}^{2}} \times \nonumber\\
&&	\left[v^r (1- c_{s}^{2}) 
	- \sqrt{c_{s}^{2} (1-v^2)[g^{rr}(1-v^2c_{s}^{2})
							-v^r v^r (1-c_{s}^{2})]
							}	\right] \nonumber
\end{eqnarray}

\noindent and their corresponding linearly independent eigenvectors:

\begin{eqnarray}
{\bf r}_{1} &=&
	\left[
	\begin{array}{c}
	\frac{\kappa}{hW(\kappa - \rho_0 c_{s}^{2})}\\ 
	v_r\\ 
	1 - \frac{\kappa}{hW(\kappa - \rho_0 c_{s}^{2})}
	\end{array}
	\right], \label{eq:eigenvector1}
\end{eqnarray}

\begin{eqnarray}
{\bf r}_{2} &=&
	\left[
	\begin{array}{c}
	1\\ 
	hW\left( v_r - \frac{v^r - (\lambda_{2} + \beta^r)/\alpha}{g^{rr} - v^r(\lambda_{2} + \beta^r)/\alpha} \right)\\ 
	hW\left( \frac{\gamma^{rr} - v^r v^r}{g^{rr} - v^r(\lambda_{2} + \beta^r)/\alpha}\right) -1
	\end{array}
	\right], \label{eq:eigenvector2}
	\end{eqnarray}

\begin{eqnarray}
{\bf r}_{3} &=&
	\left[
	\begin{array}{c}
	1\\ 
	hW\left( v_r - \frac{v^r - (\lambda_{3} + \beta^r)/\alpha}{g^{rr} - v^r(\lambda_{3} + \beta^r)/\alpha} \right)\\ 
	hW\left( \frac{\gamma^{rr} - v^r v^r}{g^{rr} - v^r(\lambda_{3} + \beta^r)/\alpha}\right) -1
	\end{array}
	\right], \label{eq:eigenvector3}
\end{eqnarray}

\noindent where $v^2=g_{rr} v^r v^r$, $c_s$ is the speed of sound and $\beta^r$ is the shift vector, which in our case, according to (\ref{eq:metric}) is zero and $g_{rr}=a^2$. On the other hand, a useful property of the gas is the speed of sound, which is defined by

\begin{equation}
hc_{s}^{2}=\chi+\left(\frac{p}{\rho_{0}^{2}}\right)\kappa,~~~ \chi=\frac{\partial p }{\partial \rho_0},
~~~ \kappa = \frac{\partial p }{\partial \epsilon},
\end{equation}

\noindent where for the ideal gas equation of state (\ref{eq:eos}), $\chi=\epsilon(\Gamma-1)=p/\rho_0$ and $\kappa = \rho_0 (\Gamma-1)$. Using expression (\ref{eq:enthalpy}) for the enthalpy, $h=1+\epsilon+p/\rho_0=1+\frac{p}{\rho_0}\frac{\Gamma}{\Gamma-1}$, one obtains an expression for the speed of sound in terms of the thermodynamical variables

\begin{equation}\label{eq:sound}
c_{s}^{2}=\frac{p\Gamma(\Gamma-1)}{p\Gamma +\rho(\Gamma-1)},
\end{equation}

\noindent which can be used to calculate the eigenvalues and eigenvectors in (\ref{eq:eigenvalues}) and (\ref{eq:eigenvector1},\ref{eq:eigenvector2},\ref{eq:eigenvector3}).


\subsection{Numerical methods}


\subsubsection{Finite Volumes} 

\begin{figure}[htp]
\includegraphics[width=8cm]{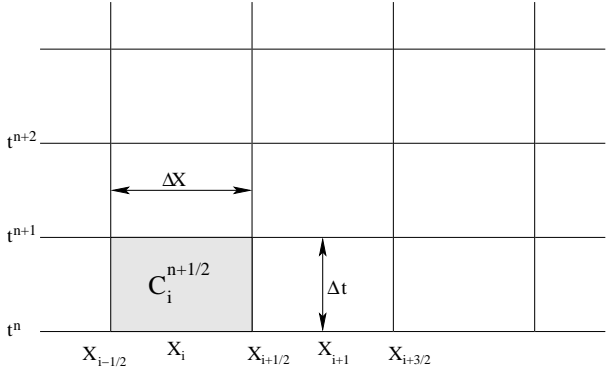}
\caption{\label{fig:FiniteVol}  In this  figure, we present the discretization and cell structure of the space-time.  Here, the center of cell $C_i^{n+1/2}$ is located at $(t^{n+1/2},x_i)$ and its space-time volume is $V=\Delta t \Delta x$. }
\end{figure}

Due to the non-linearity of the relativistic Euler equations, the presence of discontinuities (shocks) in the hydrodynamical variables is common, even if smooth initial data are considered.  For this reason, numerical methods based in the continuity of the functions, like finite differences, are not suitable to solve this kind of non-linear equations. One of the most widely used approaches in the literature is the finite volume method; firstly this method considers the problem is defined on a mesh of grid points that define a cell structure on the space-time, that is, the time is restricted to have the discrete set of values $t^n = n  \Delta t$ and the space is considered to be discretized with cells whose spatial centers are defined at $x_i = i \Delta x$ as shown in Fig. \ref{fig:FiniteVol}. Secondly, the numerical methods for balance flux type of equations consist in discretizing the equations in their integral form. This is a method particularly useful when discontinuities are expected to appear as in the present case \cite{LeVeque}.


In order to illustrate the finite volume method, consider the following conservative system of equations in one spatial dimension:

\begin{equation}\label{eq:examcons}
\partial_t {\bf u} + \partial_x{\bf F}({\bf u}) = {\bf S},
\end{equation}

\noindent where ${\bf u}$ is the vector of conservative variables, ${\bf F}({\bf u})$ are fluxes which depend on the variables ${\bf u}$ and ${\bf S}$ is a vector of sources. In this way, the first step to discretize the integral form of the equation (\ref{eq:examcons}) is to take its average over a space-time cell  $C_i^{n+1/2}$:

\begin{eqnarray} \label{eq:average}
\nonumber && \frac{1}{\Delta t \Delta x} \int_{x_{i-1/2}}^{x_{i+1/2}} \int_{t^{n}}^{t^{n+1}} \partial_t {\bf u} dxdt + \\ \nonumber
&& \frac{1}{\Delta t \Delta x} \int_{x_{i-1/2}}^{x_{i+1/2}} \int_{t^{n}}^{t^{n+1}} \partial_x{\bf F}({\bf u})dxdt  = \\ 
&& \frac{1}{\Delta t \Delta x} \int_{x_{i-1/2}}^{x_{i+1/2}} \int_{t^{n}}^{t^{n+1}} {\bf S}dxdt ,
\end{eqnarray}

\noindent where the volume of the cell  $C_i^{n+1/2}$ is $(t^n,t^{n+1})\times(x_{i-1/2},x_{i+1/2})$. Now, as second and final step, by using Gauss' theorem, this last equation can be integrated to obtain a  discretized version of the integral form of  the system of equations (\ref{eq:examcons}): 

\begin{equation} \label{eq:discretization}
\bar{{\bf u }}_{i}^{n+1}  = \bar{{\bf u}}_{i}^{n}  - \frac{\Delta t}{\Delta x}(\bar{{\bf F}}_{i+1/2}^{n+1/2} - \bar{{\bf F}}_{i-1/2}^{n+1/2}) + \bar{{\bf S}}_{i}^{n+1/2}  \Delta t  
\end{equation}

\noindent where $\bar{{\bf u}}_{i}^{n}$ are the spatial averages of the conservative variables,  $\bar{{\bf F}}_{i+1/2}^{n+1/2}$ are the temporal averages of the fluxes and $\bar{{\bf S}}_{i}^{n+1/2}$ is the spatial and temporal average of the sources. The key problem consists in the calculation of the temporal averages of the fluxes which we describe below.


\subsubsection{Cell reconstruction}

Equation (\ref{eq:discretization}) provides an evolution rule for the conservative variables, however the difficulty  to calculate the temporal averaged fluxes $\bar{{\bf F}}_{i+1/2}^{n+1/2}$ at interfaces between cells, still remains. One way to solve this problem defines the Godunov-type numerical methods \cite{Godunov}. The main idea of these methods is to consider a Riemann problem at each intercell boundary, which requires to approximate the spatial average of the variables $\bar{\bf u}$ at each cell  with piecewise functions $\tilde{{\bf u}}$. 

There are different ways of reconstructing the variables. The simplest reconstruction was introduced by Godunov and consists in defining the variables to be constant piecewise \cite{Godunov,LeVeque}. More general reconstructions assume the variables are linear piecewise, which is the case we handle in this paper; in this case, the variables $\tilde{{\bf u}}$ are reconstructed using a minmod slope limiter that restricts the slope of the linear functions defining the variables within each cell \cite{Toro,LeVeque}:

\begin{eqnarray}\label{eq:PWL}
&& \nonumber \tilde{{\bf u}}^{L}_{i+1/2} = \bar{{\bf u}}_{i} + {\bf \sigma}_{i}( x_{i+1/2} - x_{i}), \\
&& \tilde{{\bf u}}^{R}_{i+1/2} = \bar{{\bf u}}_{i+1} + {\bf \sigma}_{i+1}( x_{i+1/2} - x_{i+1}),
\end{eqnarray}

\noindent where $L$ and $R$ indicate the cell to the left and to the right respectively from the intercell boundary we deal with. The quantity ${\bf \sigma_i}$ is calculated as

\begin{equation}\label{eq:sigma}
{\bf \sigma}_i = {\it minmod}(m_{i-1/2},m_{i+1/2}). 
\end{equation}

\noindent The function $m_{i+1/2}$ is the derivative of the conservative variables $\bar{{\bf u}}$, centered at the cell interfaces

\begin{equation}\label{eq:slope}
m_{i+1/2} = \frac{\bar{{\bf u}}_{i+1} - \bar{{\bf u}}_{i}}{x_{i+1} - x_{i}},
\end{equation} 

\noindent and the minmod slope limiter is defined by 

\begin{equation}\label{eq:minmod}
minmod(a,b)= \left\{
               \begin{array}{ll}
                 a  \quad if \quad |a| < |b| \quad and \quad ab > 0\\
                 b  \quad if \quad |a| > |b| \quad and \quad ab > 0.\\
                 0  \quad if \quad ab < 0 \\ 
               \end{array}
             \right.
\end{equation}

\begin{figure}[htp]
\includegraphics[width=8cm]{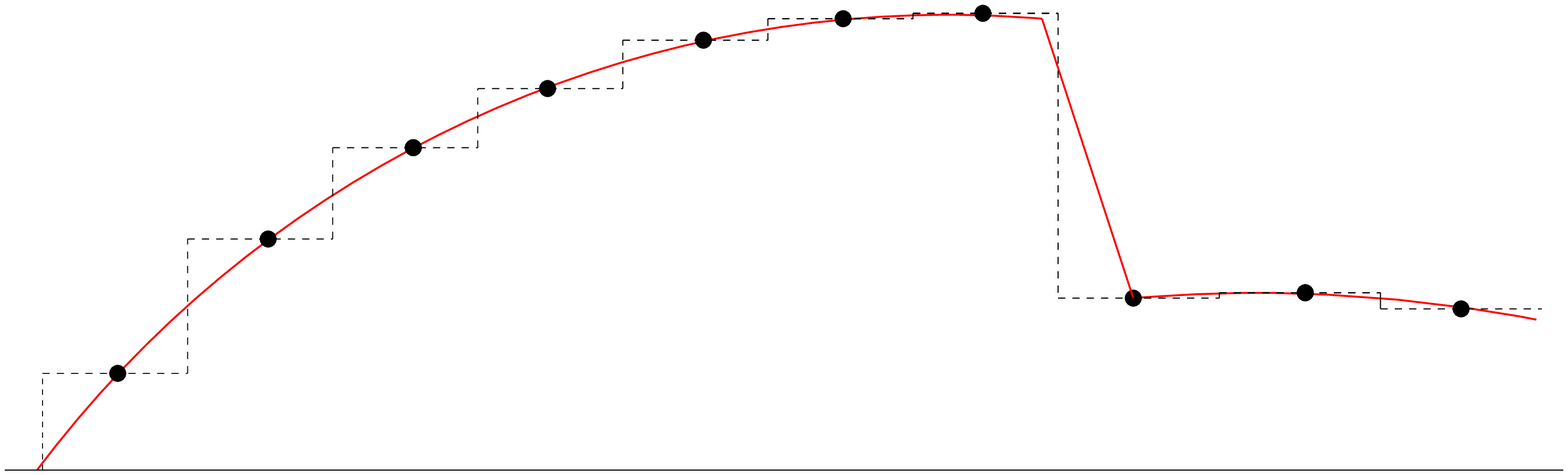}
\includegraphics[width=8cm]{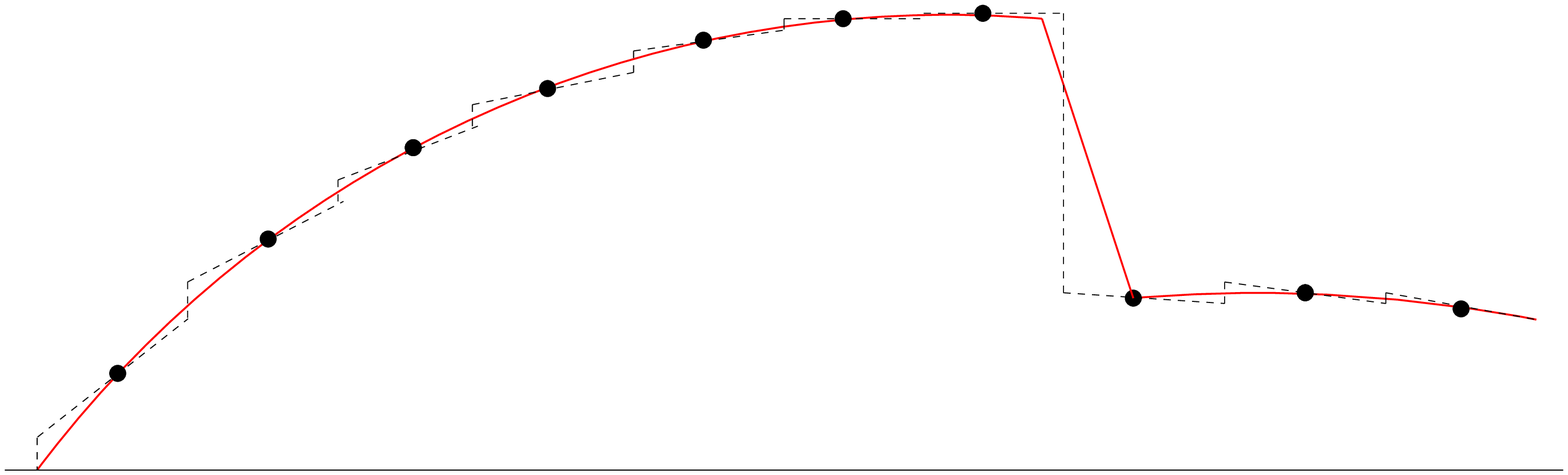}
\caption{\label{fig:minmod}  We show how a function is approximated within each cell using Godunov's constant piecewise reconstruction and the minmod linear piecewise reconstruction.}
\end{figure}

The illustration of these reconstructions are shown in Fig. \ref{fig:minmod}. The constant piecewise reconstruction provides a first order approximation in space of the variables whereas the linear reconstruction is second order accurate. This property impacts on both accuracy and order of convergence of the evolution algorithm. In this paper of course we use the linear reconstruction which helps at achieving convergence of our results. Finally, once we know the functions (\ref{eq:PWL}) to the right and left of the interfaces between cells, we can compute the temporal averaged fluxes using an approximate Riemann solver.

\subsubsection{Approximate Riemann Solver: HLLE}

Different approximate Riemann solvers require different characteristic information from the Jacobian matrix, for instance the Roe solver (see e.g. \cite{Toro}) requires the eigenvalues and the eigenvectors. One of the appealing properties of the HLLE approximate flux formula is that it requires only the eigenvalues of the Jacobian matrix. The HLLE numerical fluxes formula at the intercell boundaries reads \cite{hlle}:

{\small
\begin{equation}
\bar{{\bf F}}^{HLLE}_{i+1/2} = \frac{
\lambda^{+} {\bf F}(\tilde{{\bf u}}^L_{i+1/2}) - \lambda^{-}{\bf F}(\tilde{{\bf u}}^R_{i+1/2})
+ \lambda^{+} \lambda^{-} ( \tilde{{\bf u}}^R_{i+1/2} - \tilde{{\bf u}}^L_{i+1/2})
}
{\lambda^{+} - \lambda^{-}}\label{eq:hlle}
\end{equation}
}

\noindent where the different $\lambda$s are defined by

\begin{eqnarray}
\lambda^{+} &=& \max(0,\lambda_1^R,\lambda_2^R,\lambda_3^R,\lambda_1^L,\lambda_2^L,\lambda_3^L), \nonumber\\
\lambda^{-} &=& \min(0,\lambda_1^R,\lambda_2^R,\lambda_3^R,\lambda_1^L,\lambda_2^L,\lambda_3^L).\label{eq:lambda_plus_min}
\end{eqnarray}

\noindent Here $\tilde{\bf u}^L$ and $\tilde{\bf u}^R$ are the values of the conservative variables reconstructed at the right and left cells from the intercell boundary respectively. Notice however that for the cases in this paper, the fluxes and sources in (\ref{eq:FluxConservative2}) depend not only on the conservative variables, but also on $p$ and $v^r$ which are primitive variables. This requires the reconstruction of these primitive variables too. Also, the different $\lambda$s depend on the speed of sound, which in turn depends on $p$ (see \ref{eq:sound}) and is needed at both the left and right cells. This is the reason why it is required to calculate the conservative and primitive variables.


\subsubsection{Calculation of primitive variables}

As mentioned above, the numerical fluxes and sources depend both on the conservative and on the primitive variables. After each time step within an evolution scheme like that in (\ref{eq:discretization}) during the evolution, one obtains new values of the conservative variables $D_i,S_r{}_{i},\tau_i$ at each cell interface $i$ across the grid, then it is required to reconstruct the primitive variables out of the conservative ones in order to account with the necessary information to calculate the numerical fluxes  and sources (\ref{eq:hlle}) for the expression (\ref{eq:conservative_split_flux}).

In order to do so, first of all we recognize that the variables that are being evolved in our system of equations are ${\bf w} = a{\bf u} = (aD,aS_r,a\tau):=(w_1,w_2,w_3)$ (see equations (\ref{eq:evolution_euler})). From definition (\ref{eq:cons_vars}) one can solve for two of the primitive variables:

\begin{eqnarray}
\rho_0 &=& \frac{D}{W} = \frac{w_1}{a} \sqrt{1-a^2 (v^r)^2},\label{eq:prim_rho}\\
v^r &=& \frac{S_r}{a^2 (\tau +p + D)} = \frac{w_2}{a^2 (w_3 + ap + w_1)},\label{eq:prim_v}
\end{eqnarray}

\noindent whereas the pressure is given by:

\begin{eqnarray}
p&=& \rho_0 \epsilon (\Gamma -1) = (\rho_0 h - \rho_0 -p)(\Gamma -1)\nonumber\\
&=&(\Gamma -1)\left[ \frac{S_r}{W^2 v_r} - \frac{D}{W} -p \right] \nonumber\\
&=& (\Gamma-1) \frac{D}{W}\left[ \frac{S_r /v_r - DW - pW^2}{DW} \right]\nonumber\\
&=& \rho_0 (\Gamma-1)\left[ \frac{\tau + D(1-W) + p(1-W^2)}{DW}\right]\nonumber\\
&=& \rho_0 (\Gamma-1) \left[ \frac{w_3 + w_2 (1-W) +ap(1-W^2)}{w_1 W} \right],\label{eq:pressure}
\end{eqnarray}

\noindent where $W=W(v^r (p))$. This defines a trascendent equation for $p$ which has to be solved at each cell in the domain. We solve this equation using a Newton-Rapson root finder for $p$ at each cell. With this it is possible to reconstruct the $p$ in order to obtain $p^L$ and $p^R$. Then using $\tilde{\bf u}^L$ and $\tilde{\bf u}^R$ together with (\ref{eq:prim_v}) we calculate $v^L$ and $v^R$ and using (\ref{eq:prim_rho}) the rest mass density. Then it is possible to calculate the speed of sound on the left and right using (\ref{eq:sound}), with this one can calculate the eigenvalues of the Jacobian matrix (\ref{eq:eigenvalues}), which in turn allows one to calculate $\lambda^{+}$ and $\lambda^{-}$ using (\ref{eq:lambda_plus_min}) and finally the numerical fluxes (\ref{eq:hlle}).

\subsection{Evolution}

The evolution algorithm can be summarized as follows:

\begin{enumerate}
\item Start with given values of the primitive variables that contain the physically relevant information of the problem.
\item Calculate the conservative variables.
\item Reconstruct all the conservative variables at the left and right from the intercell boundaries using 
 (\ref{eq:PWL}) and the pressure by solving (\ref{eq:pressure}) and then also reconstructing to the left and to the right from all the intercell boundaries. With this calculate also the velocity and rest mass density at the left and right cells.
\item Calculate the speed of sound and the eigenvalues of the Jacobian matrix and use them to calculate $\lambda^{+}$ and $\lambda^{-}$.
\item Use such result to calculate the numerical HLLE fluxes (\ref{eq:hlle}). 
\item Integrate in time the expression (\ref{eq:discretization}) for equations (\ref{eq:FluxConservative2}).
\item Calculate the primitive variables 
\item Repeat from step 3 on.
\end{enumerate}

In order to evolve the averaged conservative variables from time step $t^n$ to $t^{n+1}$ we use the discrete expression (\ref{eq:discretization}) with numerical fluxes (\ref{eq:hlle}) corresponding to equations (\ref{eq:FluxConservative2}), which we perform using the method of lines (MoL) with a third order TVD Runge-Kutta integrator.


\section{The blast wave problem} 
\label{sec:blast}

The spherical blast wave problem is a particular realization of a Riemann problem. Physically, it consists in a relativistic gas distributed into two chambers separated by a removable spherical membrane located at $r=r_0$. Initially the gas in the inner chamber has a higher density and pressure than in the outer one, and the velocity is zero everywhere. Once the membrane is removed, a shock wave moves from the region of higher pressure to the region with lower pressure. Also a rarefaction wave travels in the opposite direction. Strictly speaking, there are various waves propagating in the space, a shock wave moving outwards, a rarefaction wave moving inwards and between the shock wave and the tail of the rarefaction wave two new states are developed which are separated by a contact discontinuity. 

We present different physical situations corresponding to this problem, based on the numerical solution of the relativistic Euler's equations. Specifically, the blast wave problem is set on top of the Minkowski space-time. All we need to do is set  the metric functions to the values $\alpha=a=1$ in (\ref{eq:metric}). In order to illustrate the physics of the spherical blast wave problem, we perform different simulations. The parameters we use are shown in Table \ref{tab:BWinitialdata}. 

\begin{figure*}[htp]
\includegraphics[width=8cm]{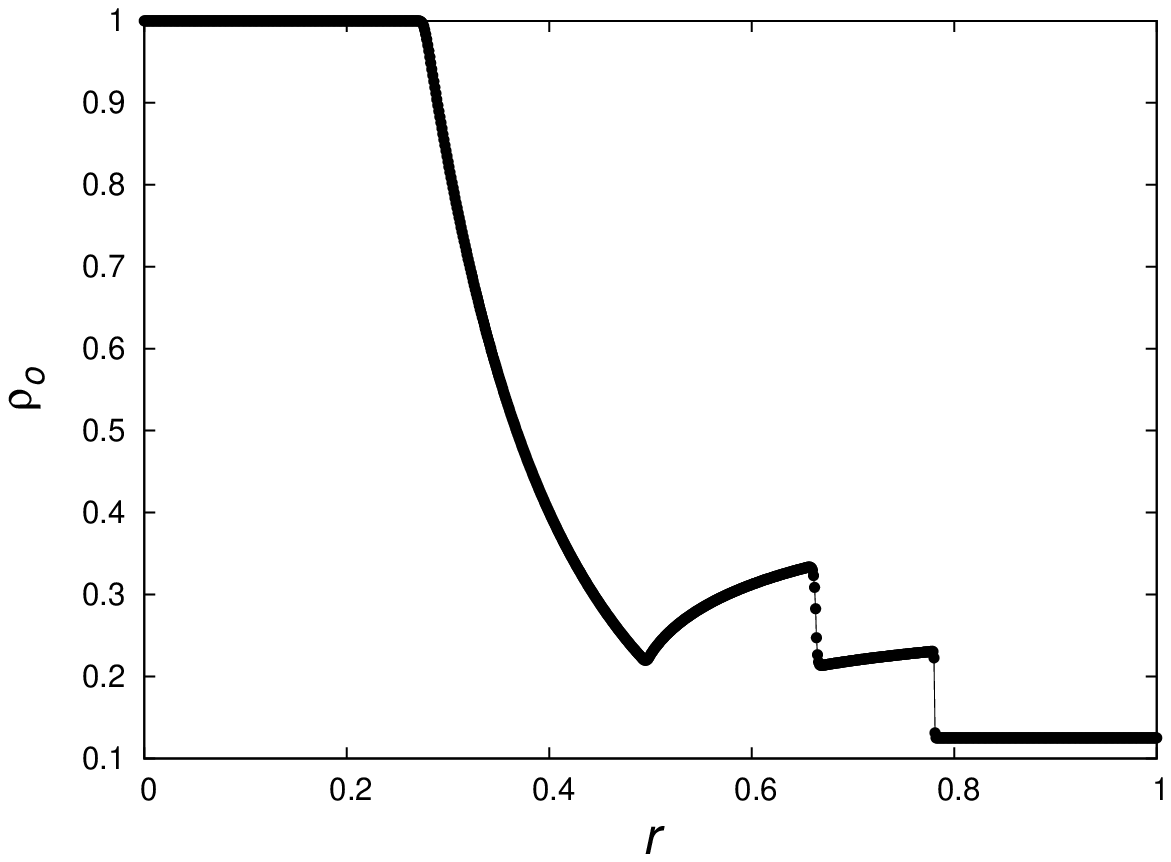}
\includegraphics[width=8cm]{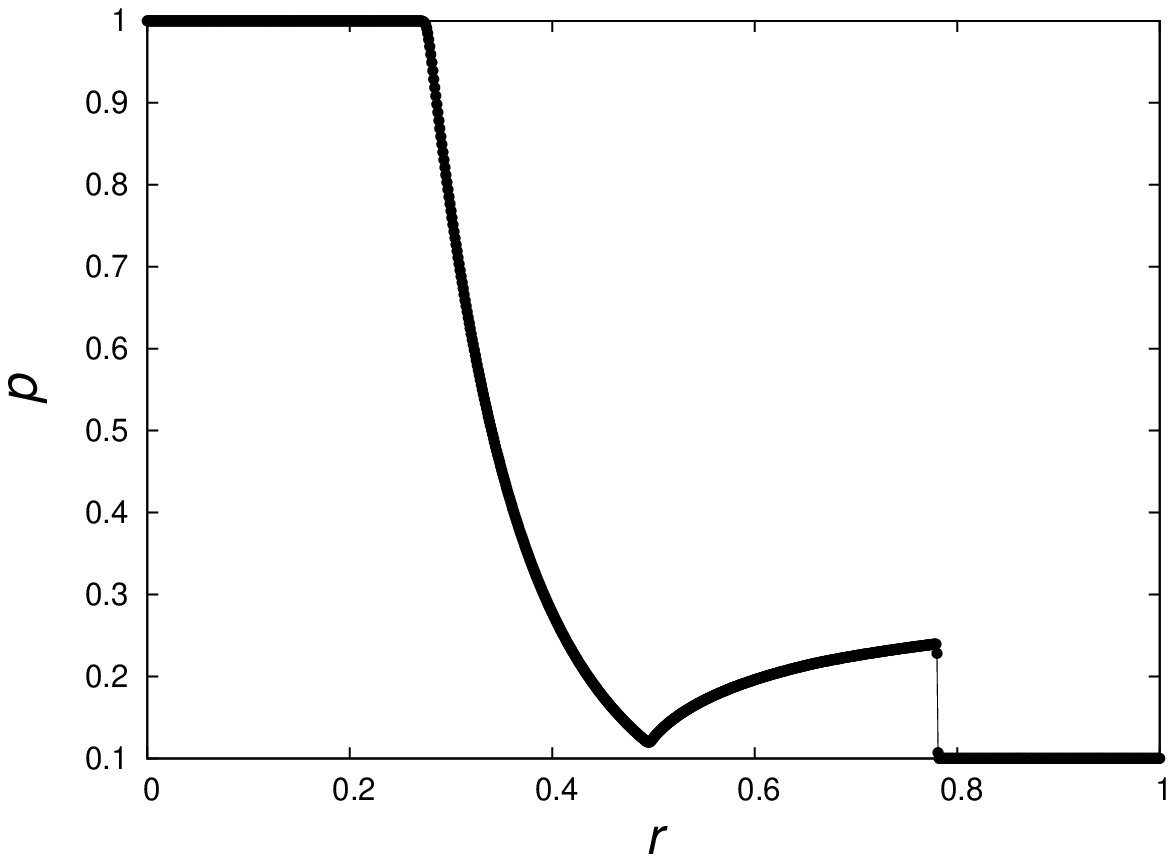}
\includegraphics[width=8cm]{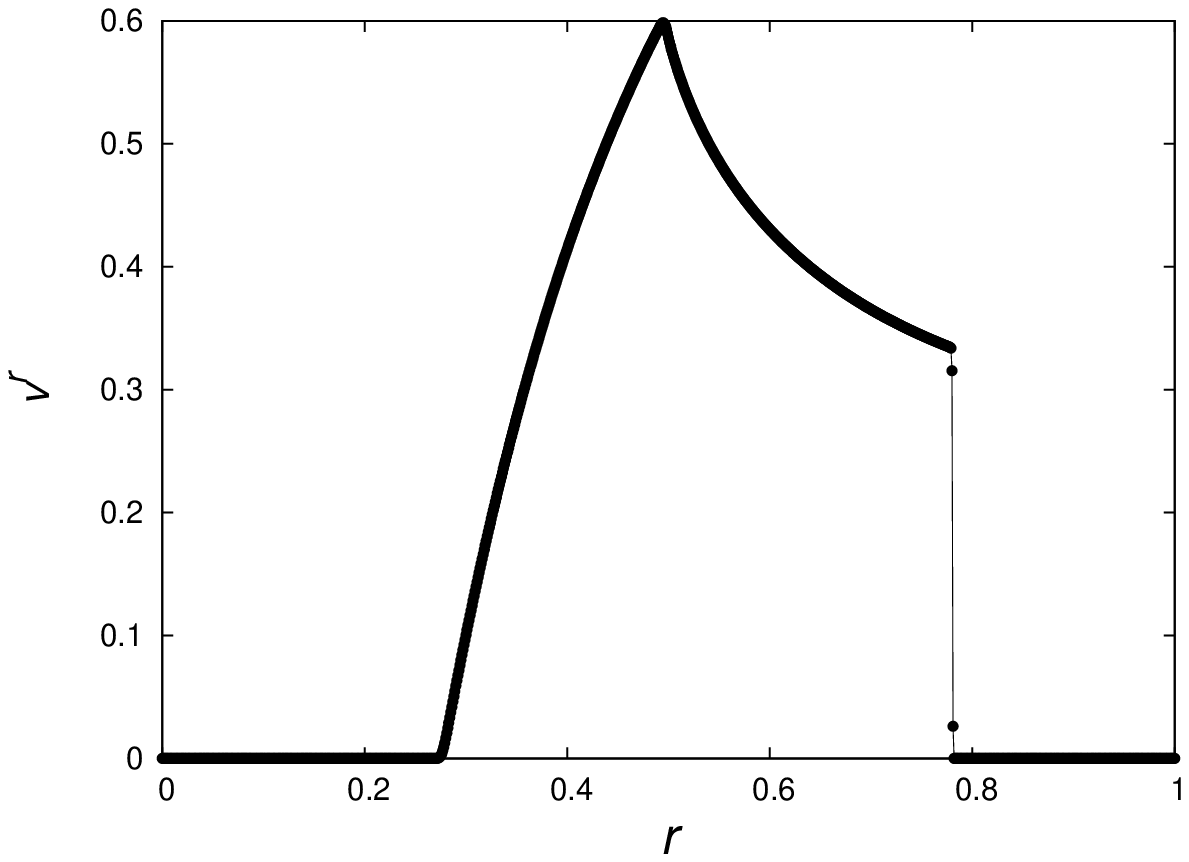}
\includegraphics[width=8cm]{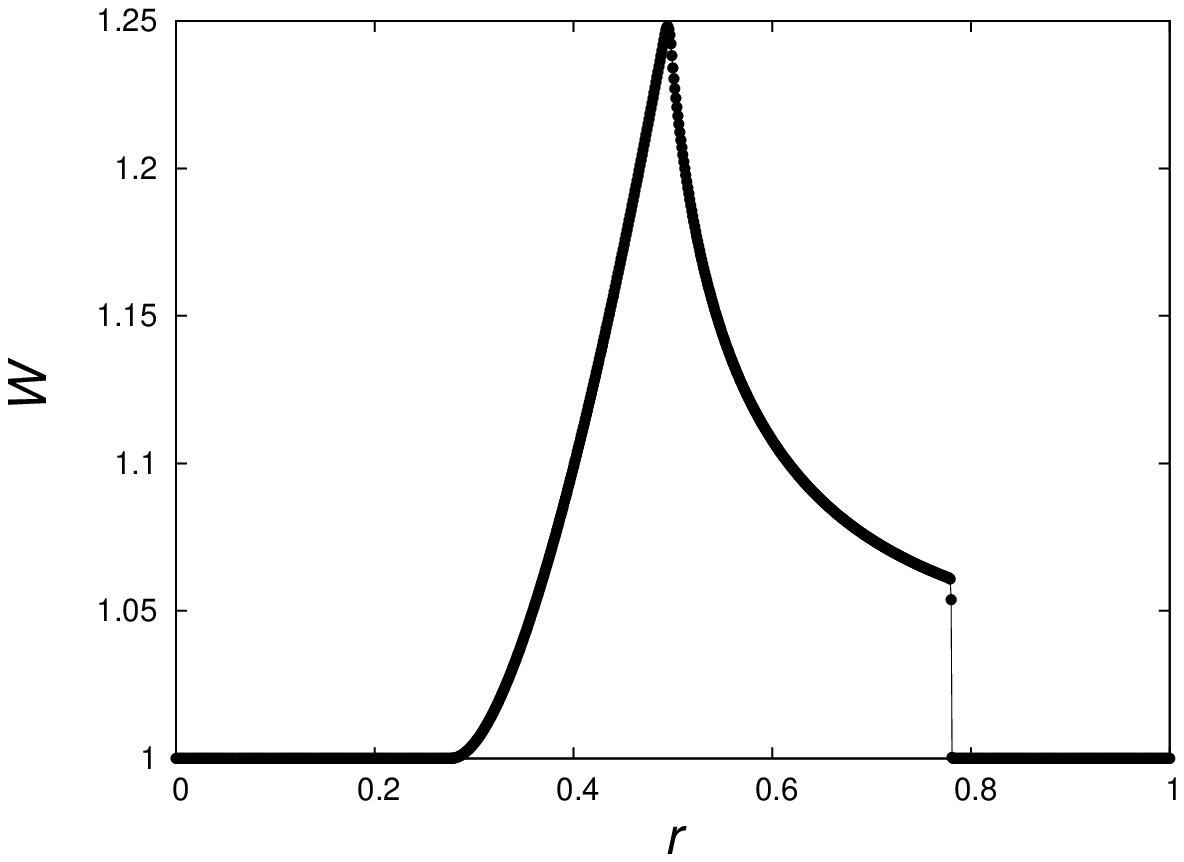} 
\includegraphics[width=8cm]{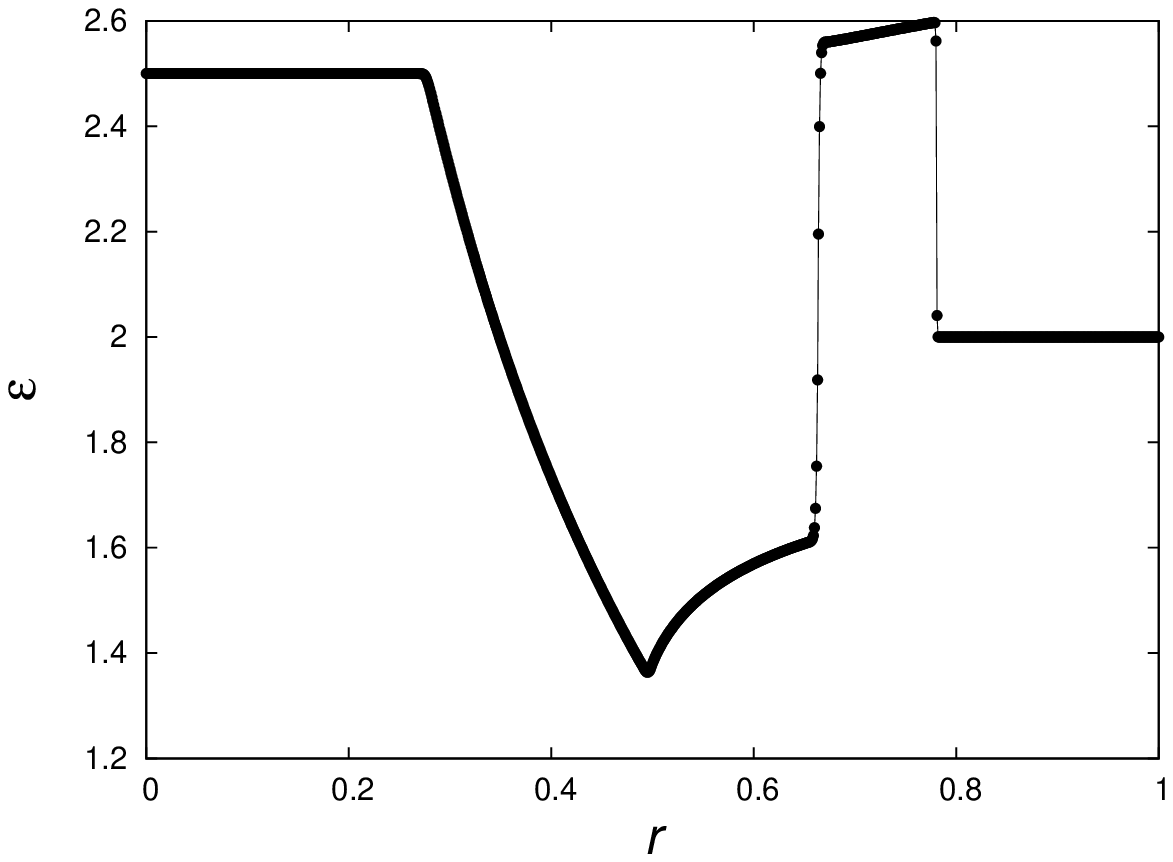}
\includegraphics[width=8cm]{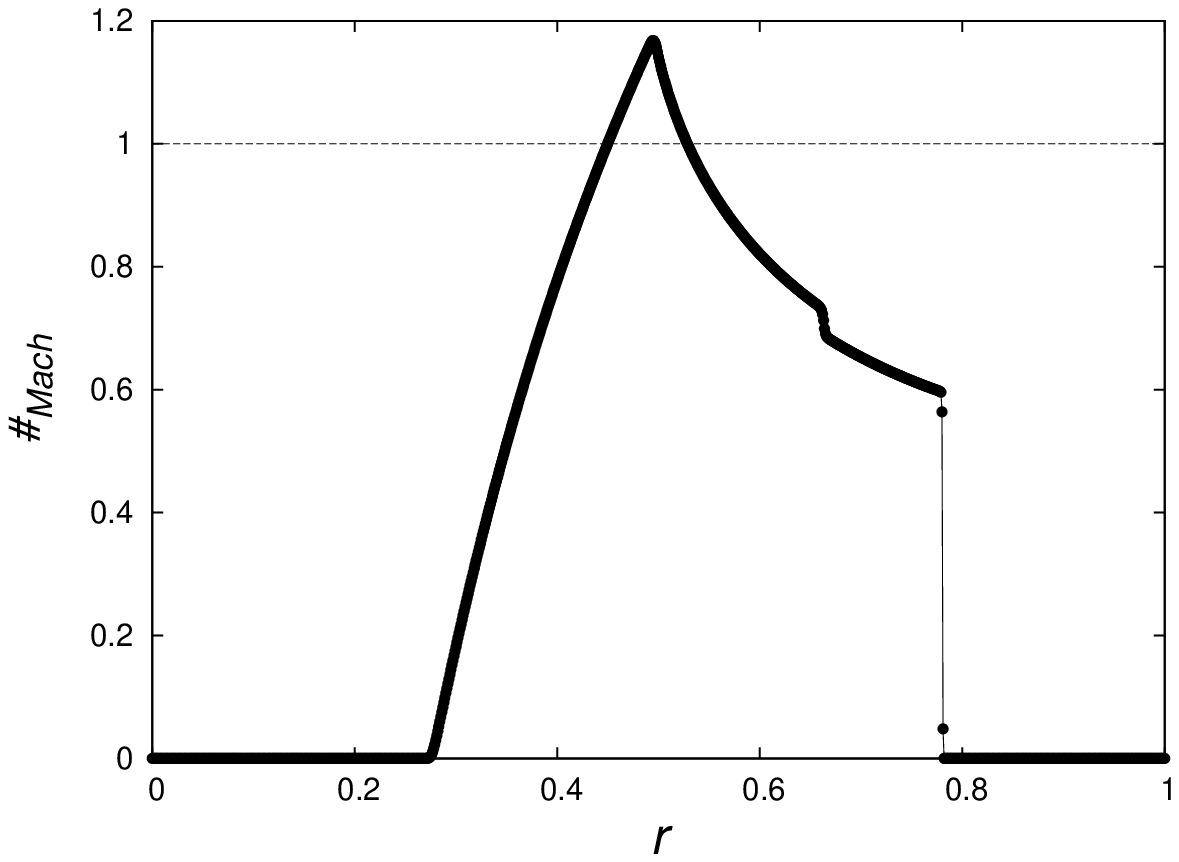} 
\caption{\label{fig:WeakBW}  Weak blast wave model at $t=0.4$ is shown. The initial discontinuity is located at $r=0.5$ and the adiabatic index  is $\Gamma=1.4$. The spatial resolution used to carry out this numerical simulation is $dr = 2\times10^{-4}$ with a Courant factor of $\Delta t / \Delta x = 0.25$. Notice that there is  small region where the Mach number is bigger than one, which indicates that the  fluid is supersonic.}
\end{figure*}

\begin{figure*}[htp]
\includegraphics[width=8cm]{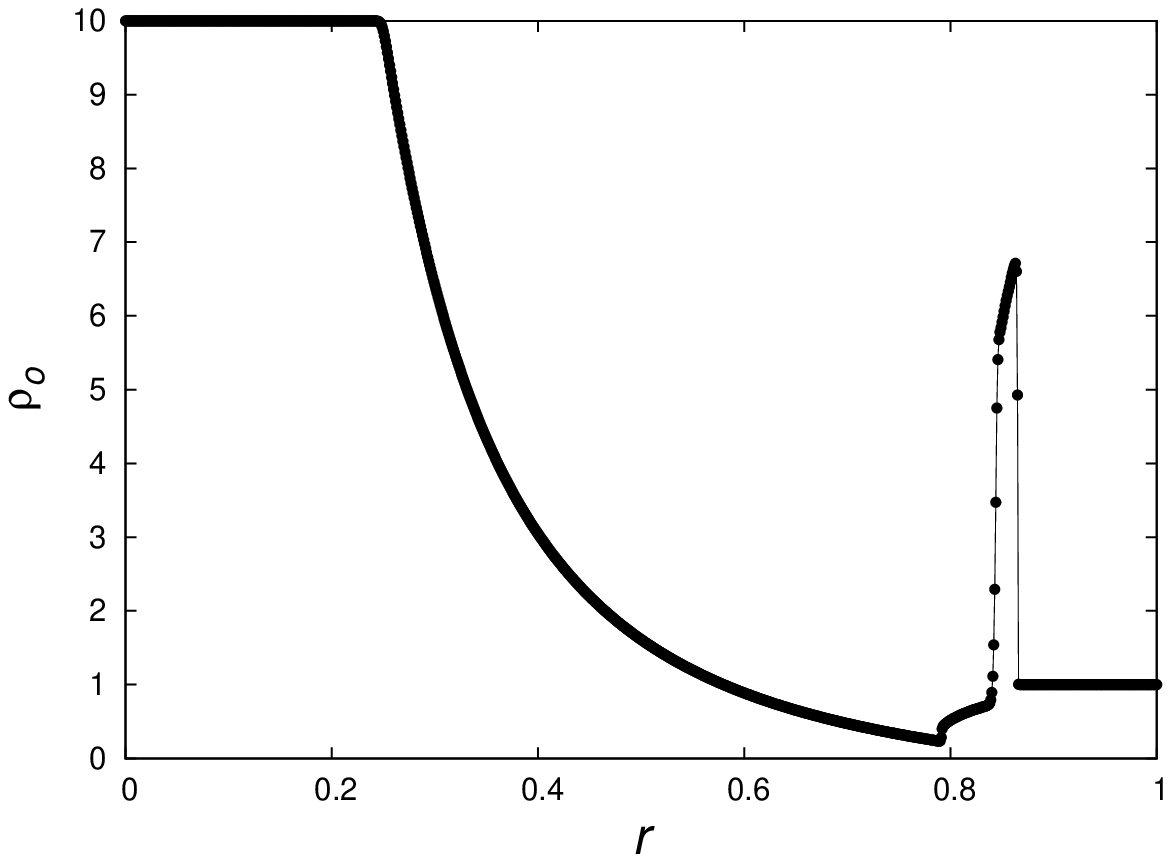}
\includegraphics[width=8cm]{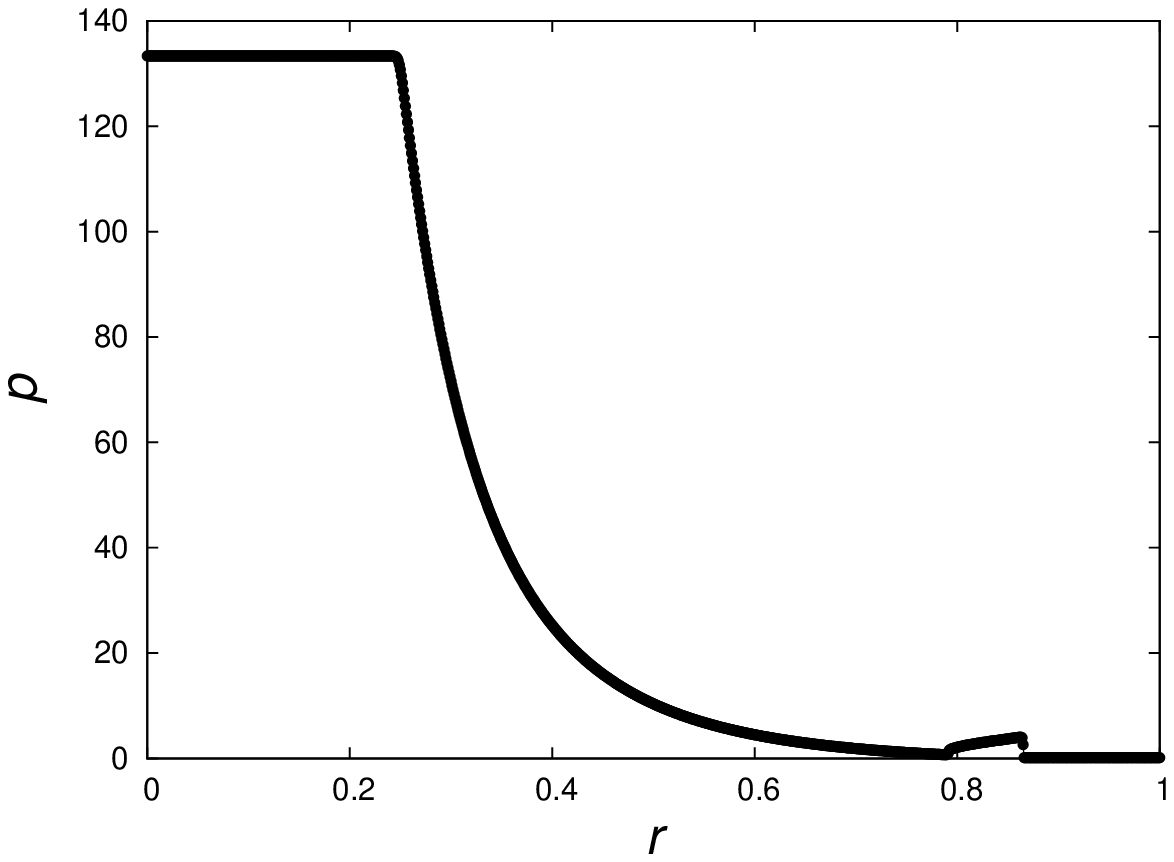}
\includegraphics[width=8cm]{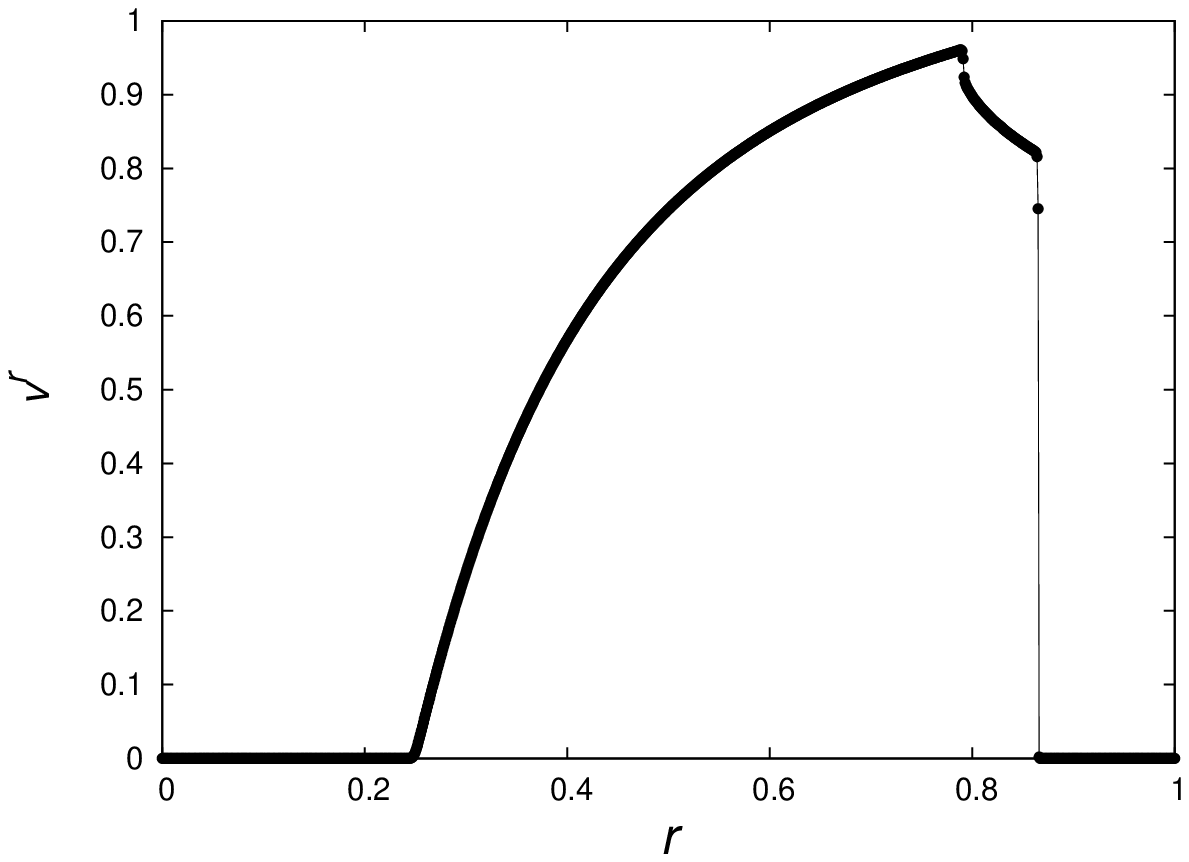}
\includegraphics[width=8cm]{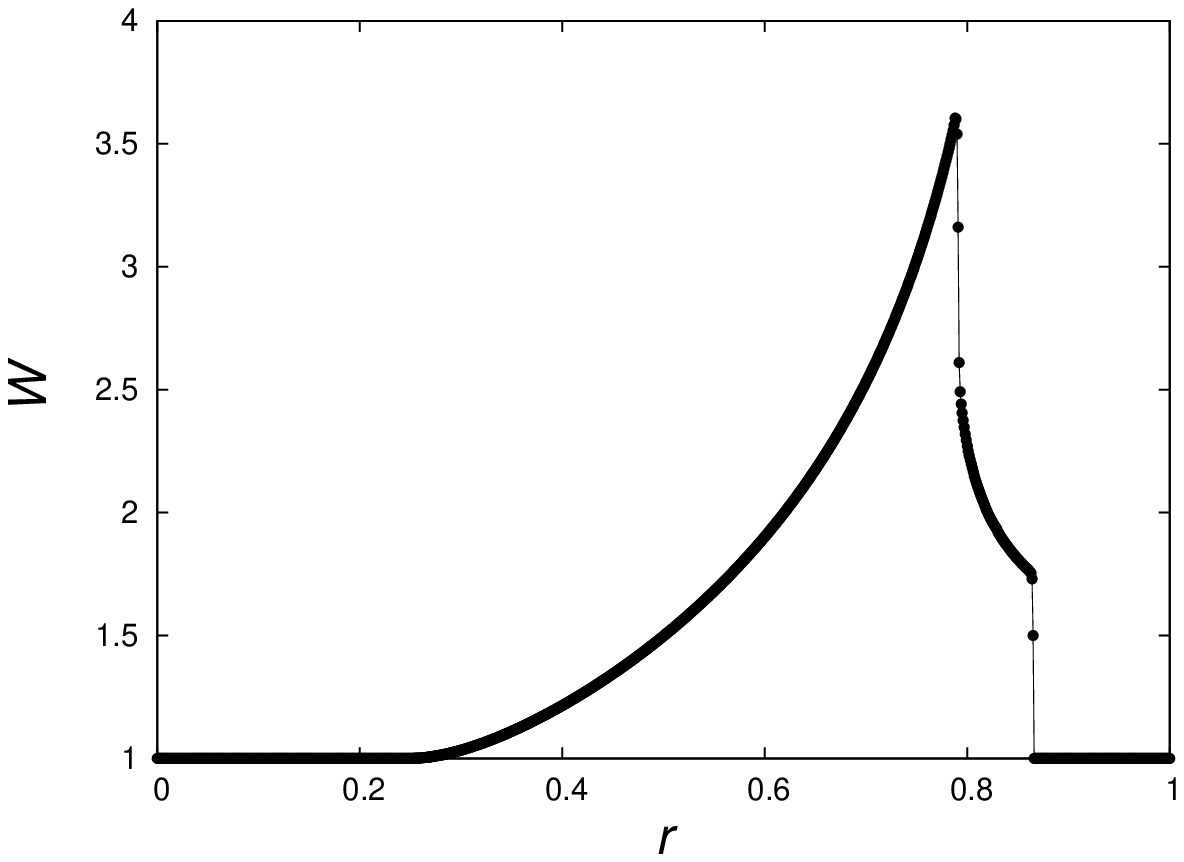} 
\includegraphics[width=8cm]{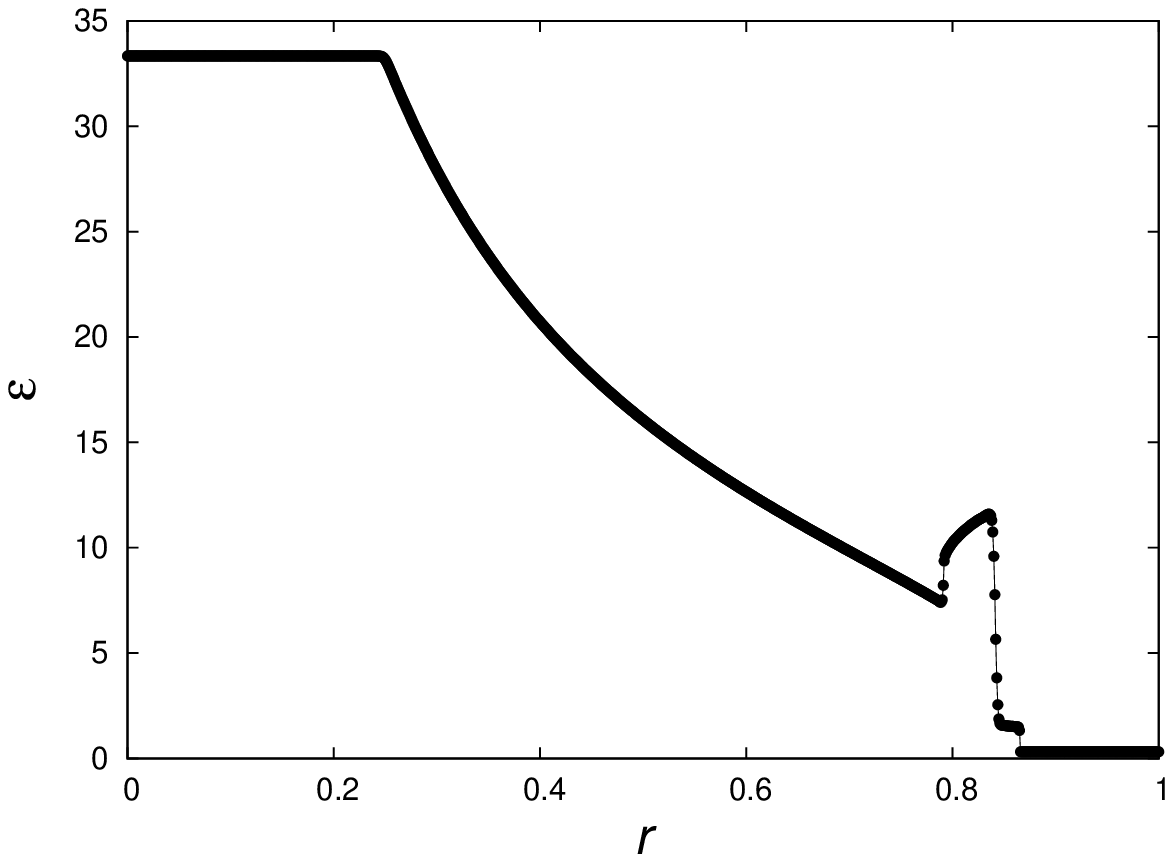}
\includegraphics[width=8cm]{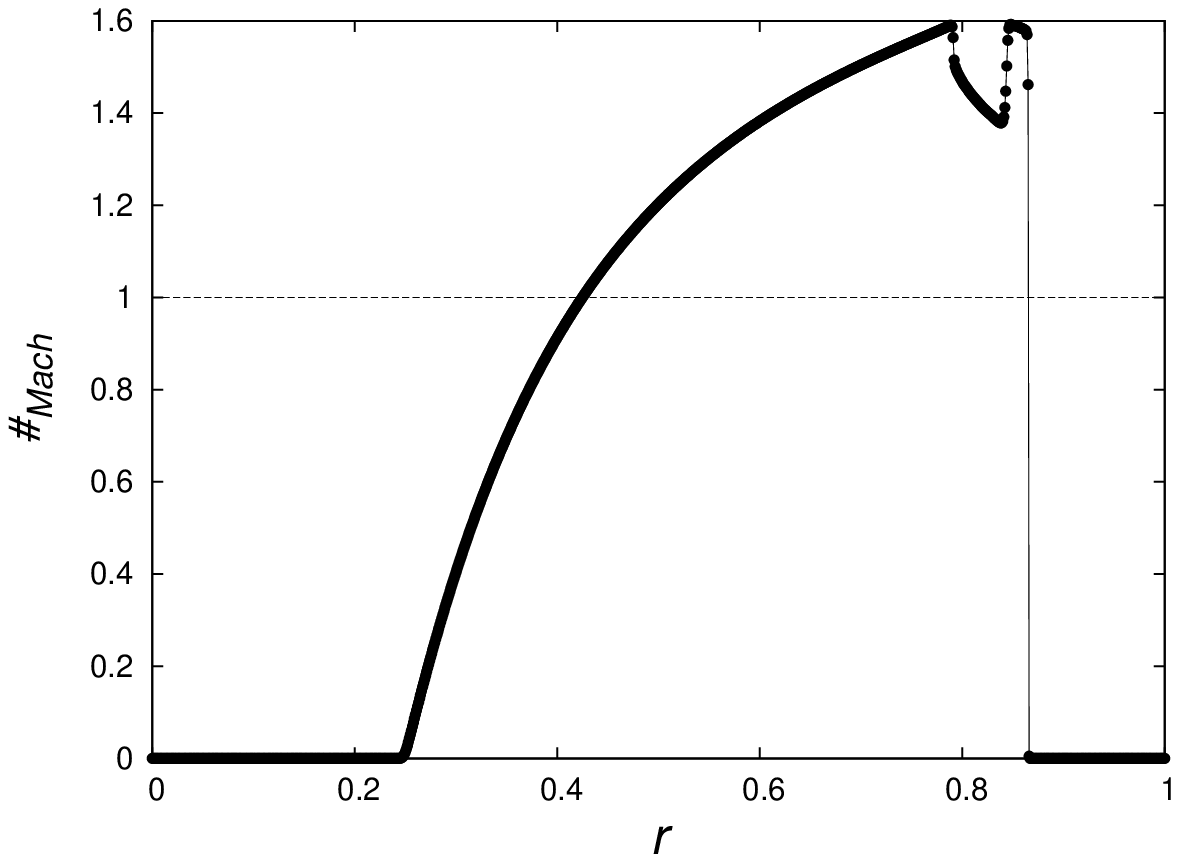} 
\caption{\label{fig:StrongBW} Strong blast wave model. The numerical parameters are as in the previous case and the gas obeys the same equation of state with $\Gamma=1.4$. In this case the Lorentz factor reaches values near 4. The region where the velocity of the fluid is supersonic is wider and the velocity there higher than in the previous case.} 
\end{figure*}

In the weak blast case shown in Fig. \ref{fig:WeakBW},  the difference of pressure is one order of magnitude, whereas in the strong blast wave case case in Fig. \ref{fig:StrongBW}, the difference of pressure in the initial shock is of three orders of magnitude, respectively. As we can see, the presence of a blast wave is more remarkable  when the difference of pressure is higher, producing velocities close to the speed of light  and regions where the fluid is supersonic. The case corresponding to the strong blast wave, has regions where the Lorentz factor approaches the value of $4$, which indicates the relativistic nature of the process. 

\begin{figure*}[htp]
\includegraphics[width=8cm]{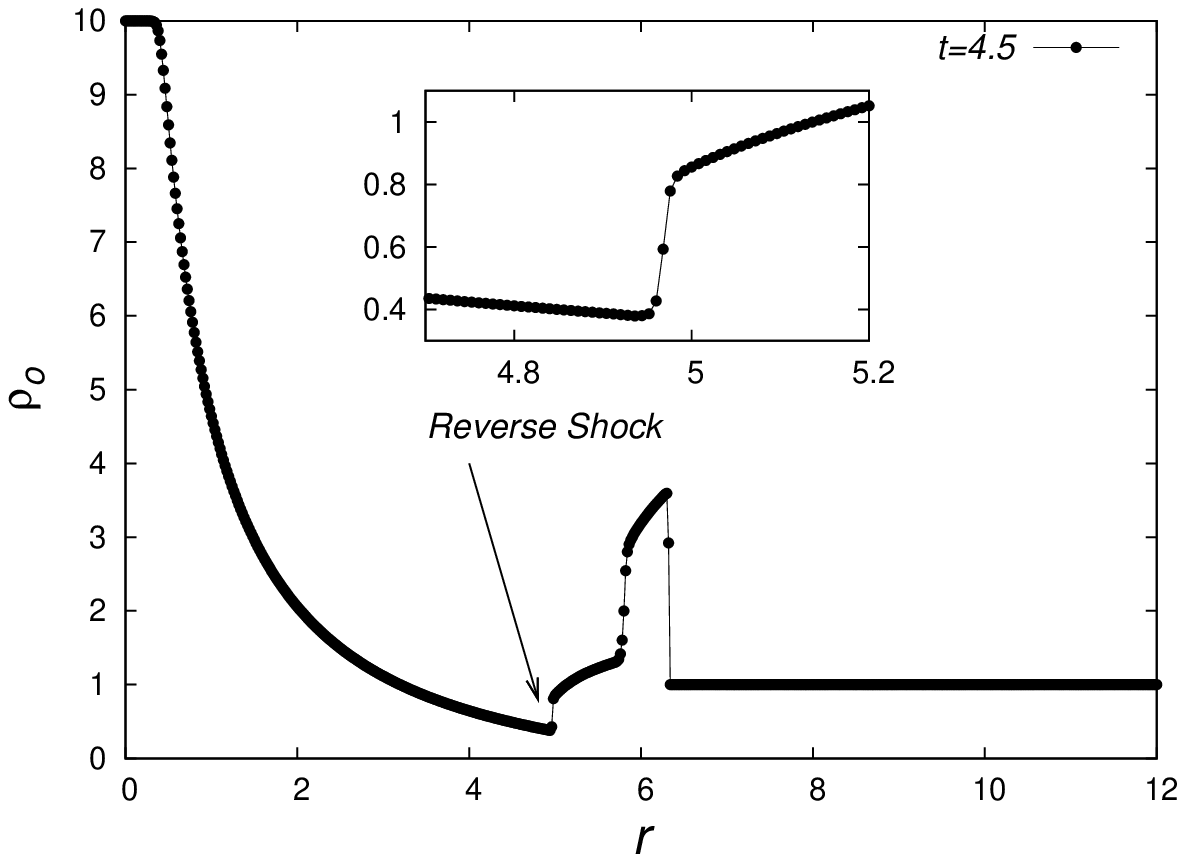}
\includegraphics[width=8cm]{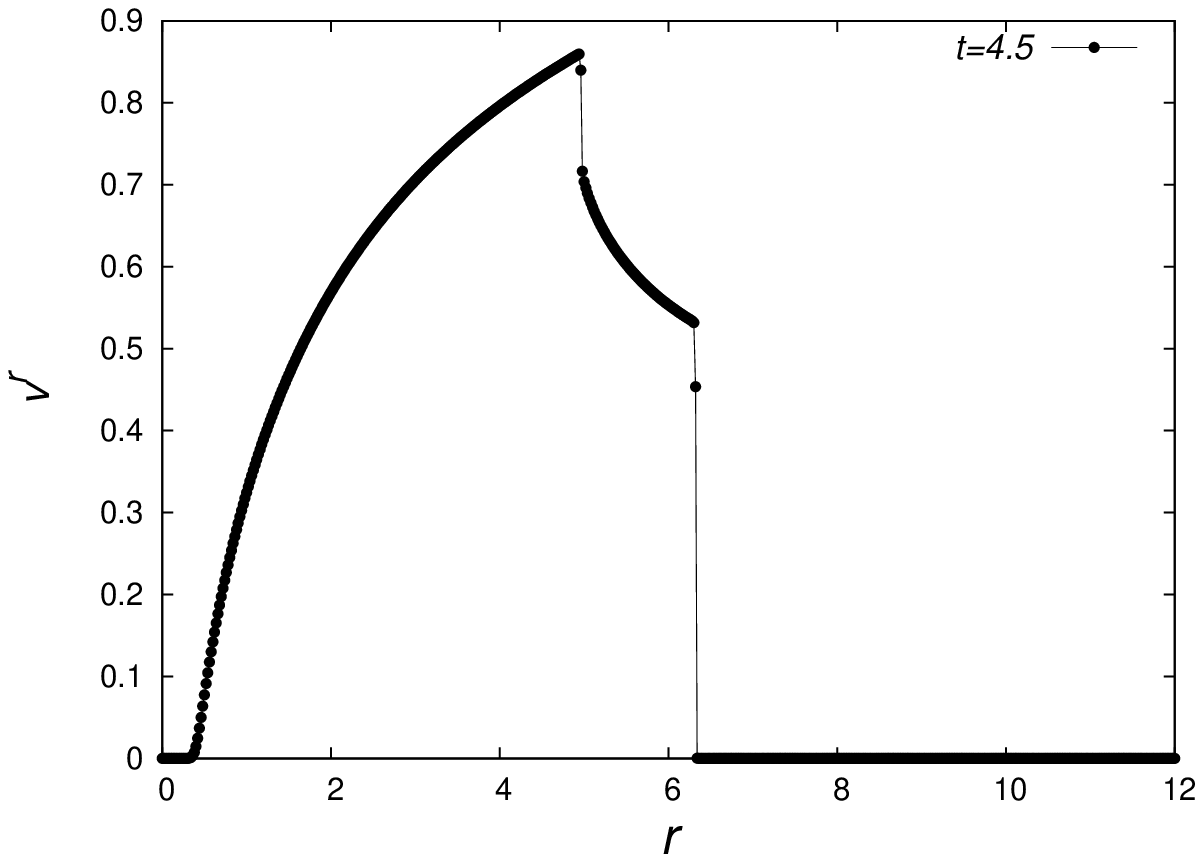}
\includegraphics[width=8cm]{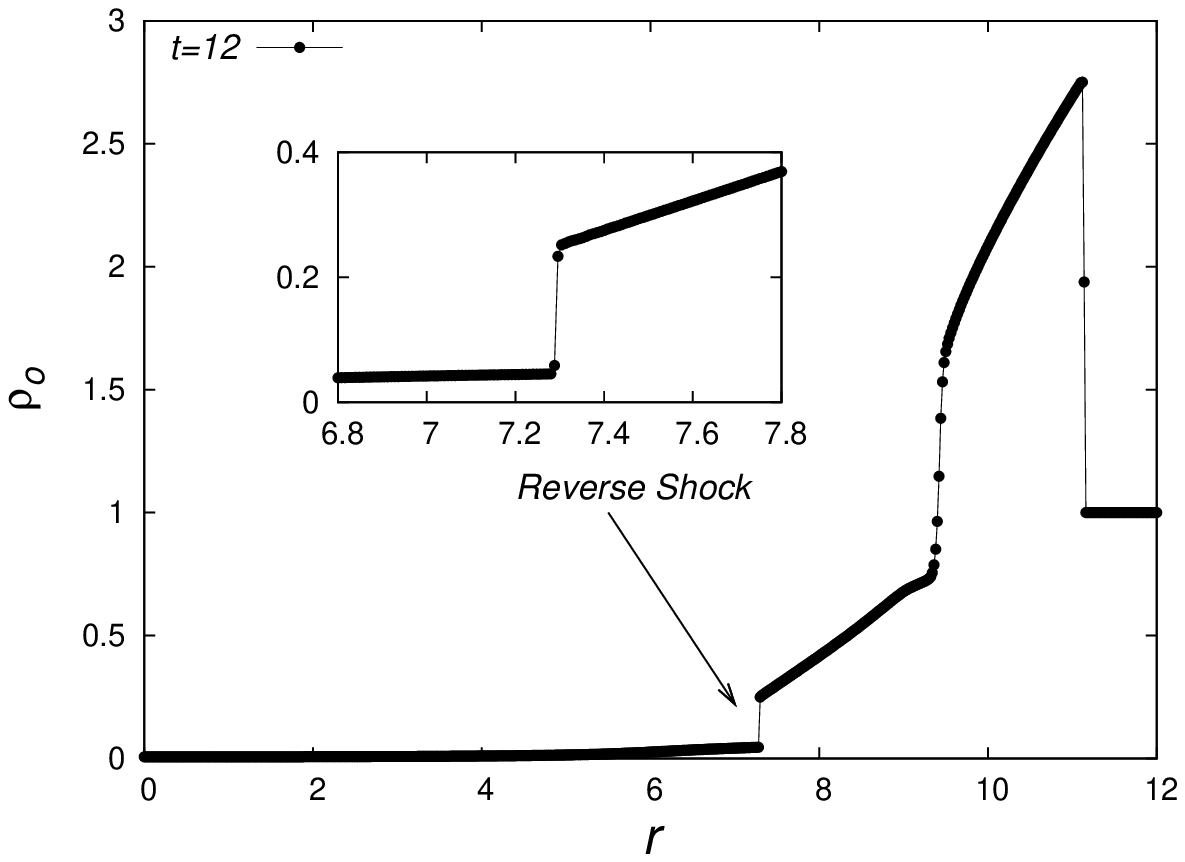}
\includegraphics[width=8cm]{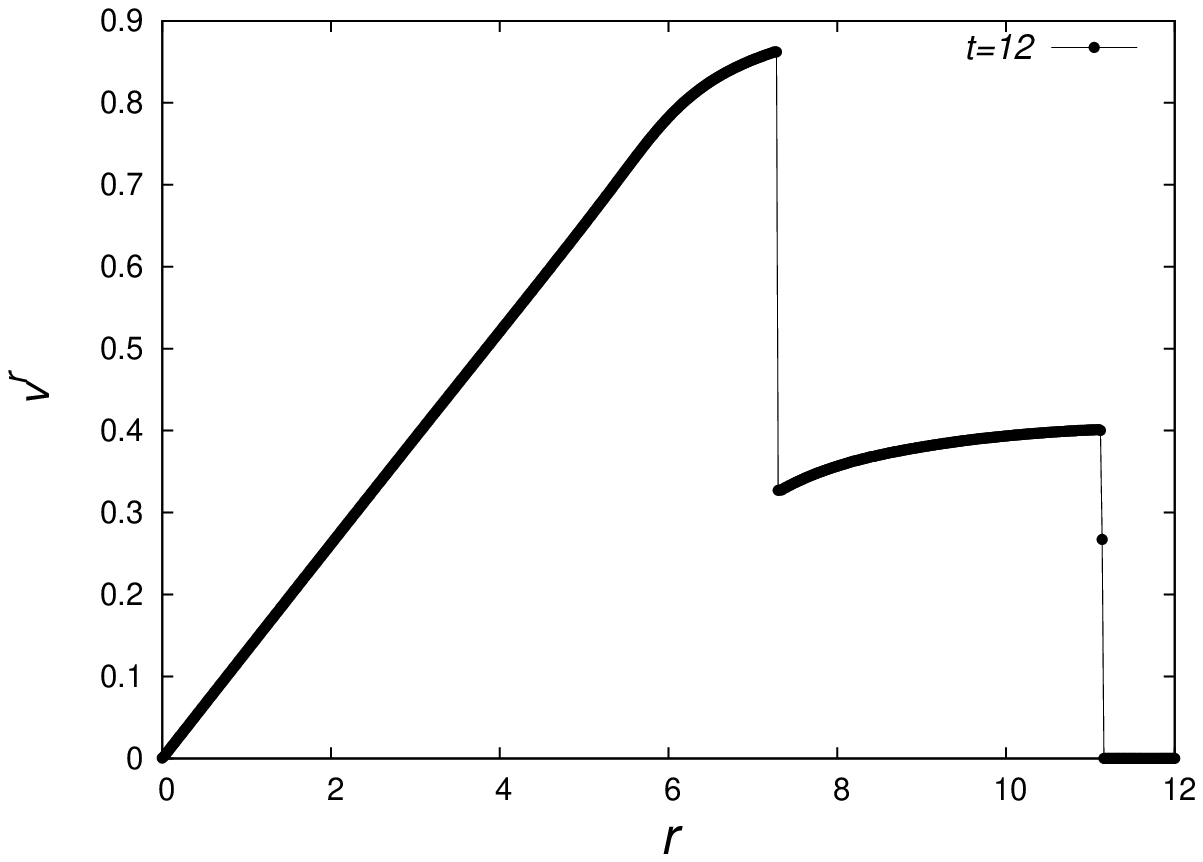}
\includegraphics[width=8cm]{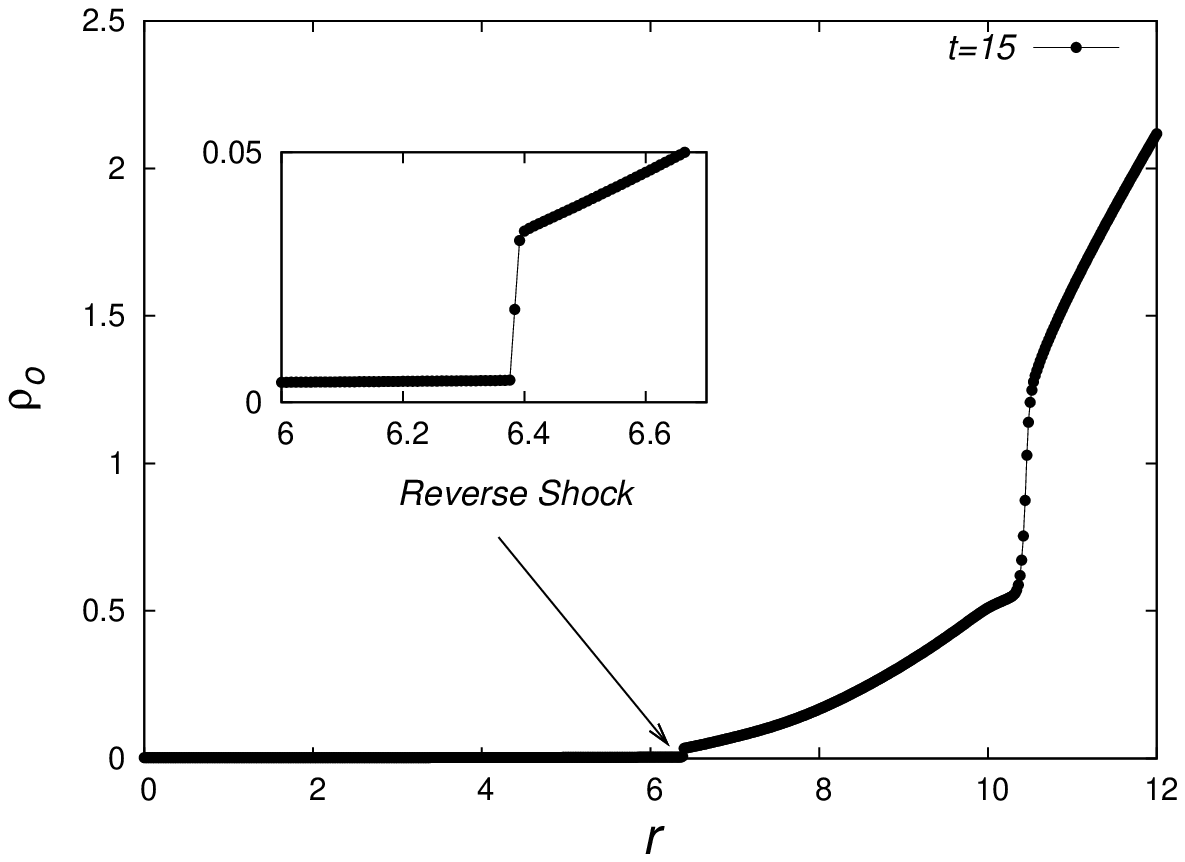}
\includegraphics[width=8cm]{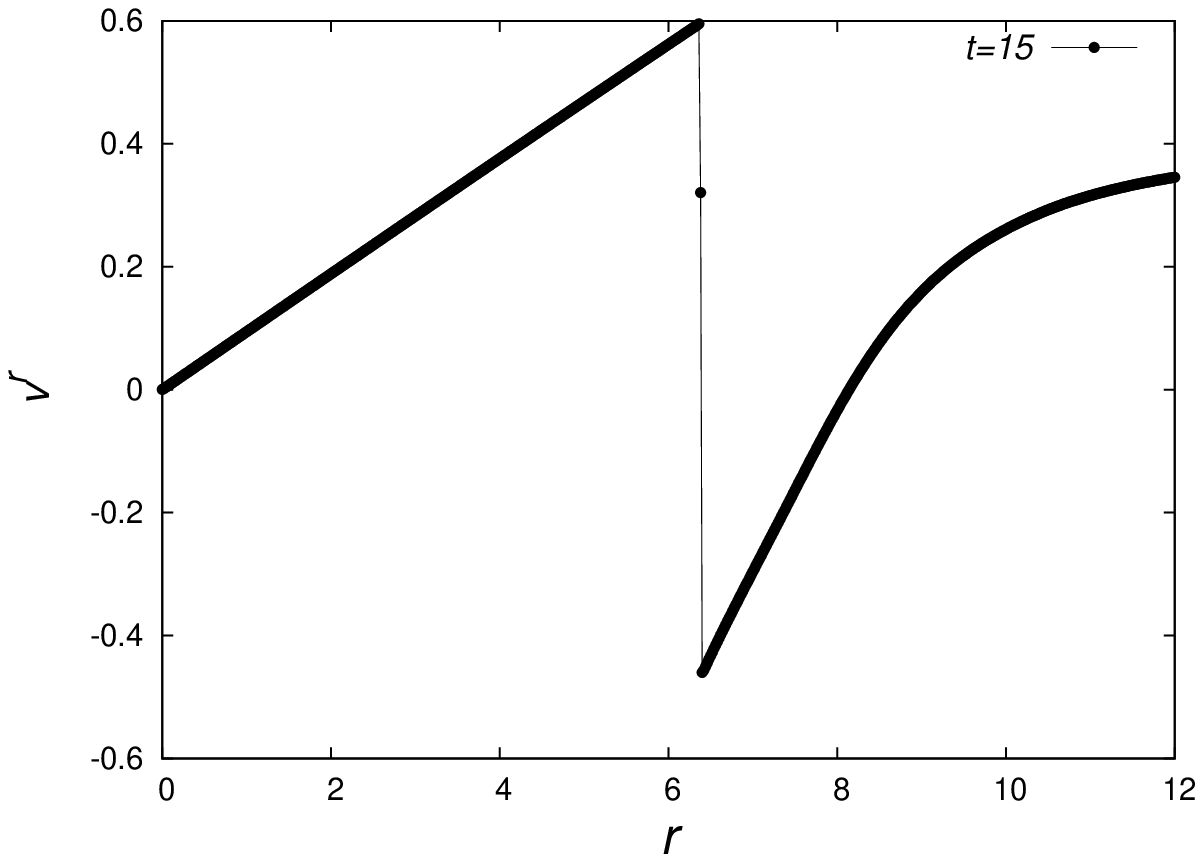}
\includegraphics[width=8cm]{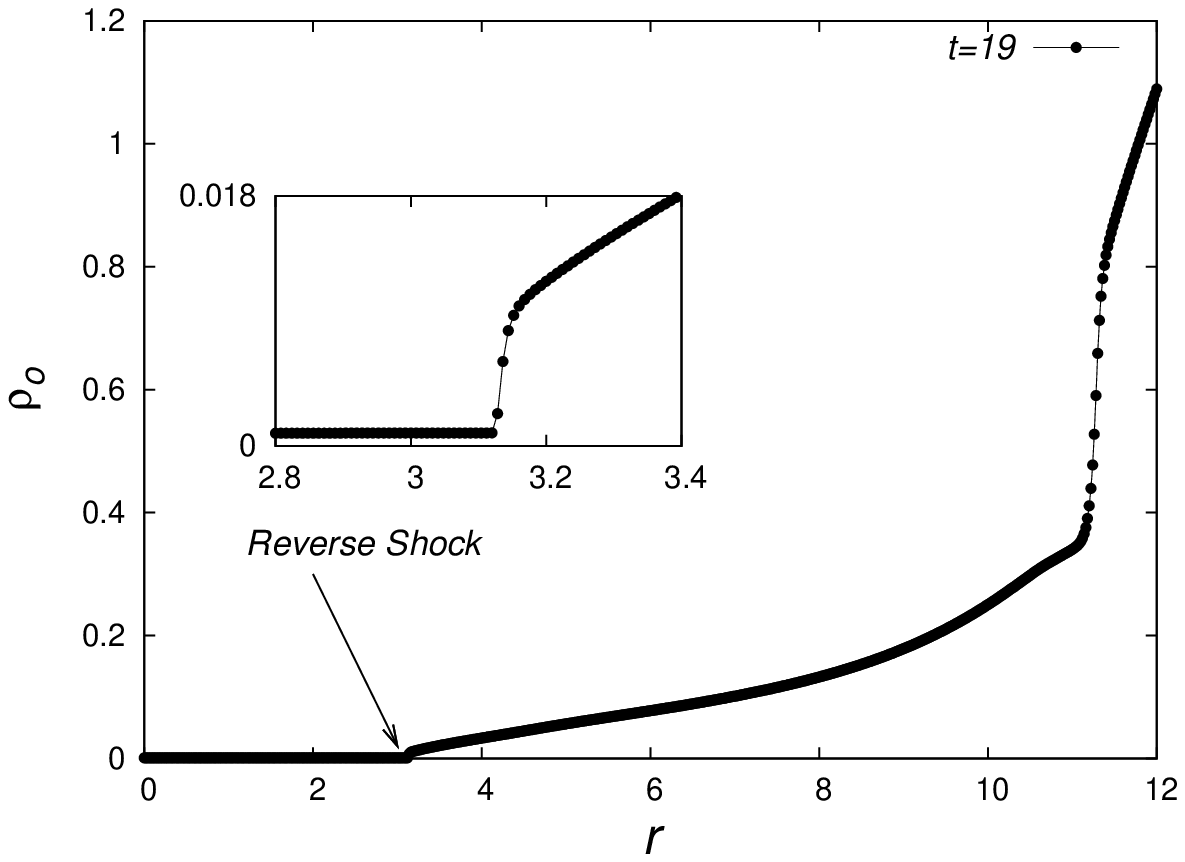}
\includegraphics[width=8cm]{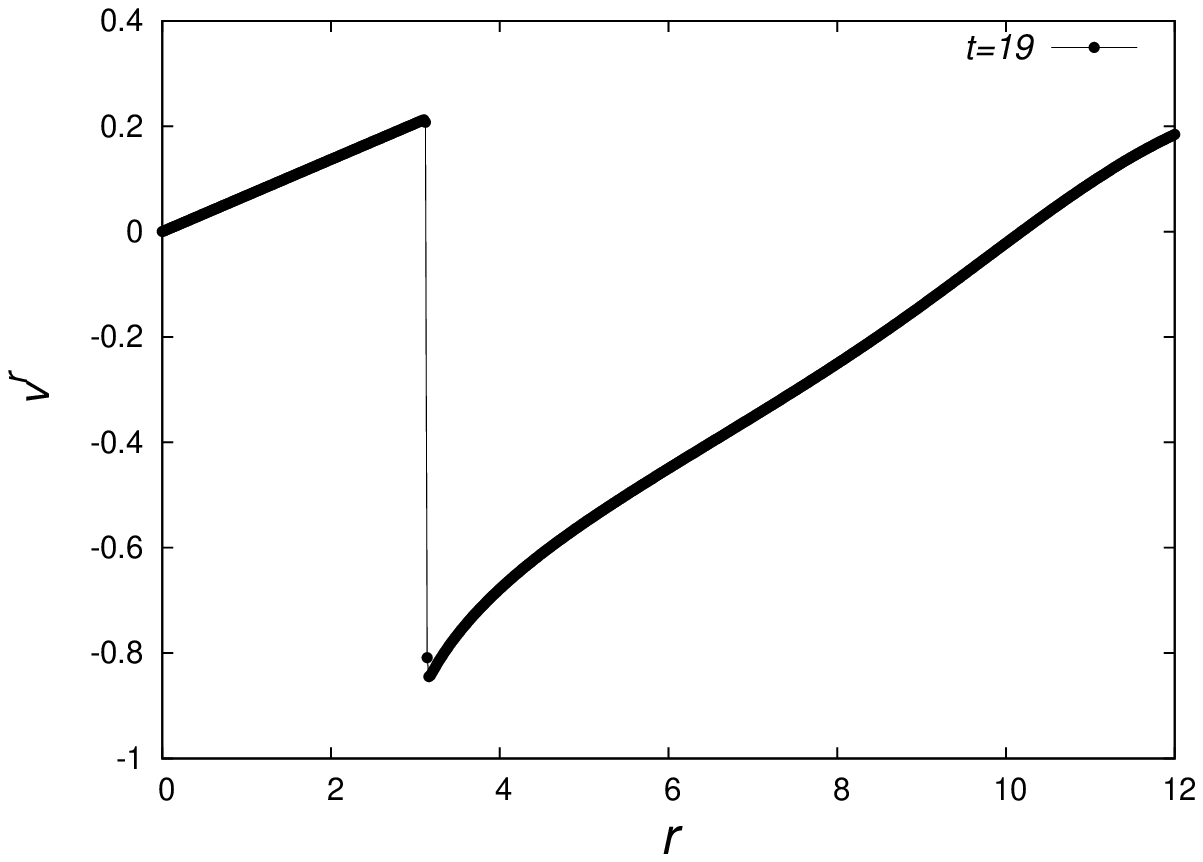}
\caption{\label{fig:Recoil} In this figure, the presence of a reverse shock wave, in the spherical blast wave problem, is shown. The parameters we use, in this example, are the following: domain of the simulation $r \in [0,12]$, numerical resolution $dr=0.004$, Courant Factor $0.25$, adiabatic index $1.4$, initial discotinuity $r_0=3.0$ and the initial primitive variables are $p_i = 13.33$, $p_e = 0.1$, $\rho_i=10$, $\rho_e=1$ and $v_i=v_e=0$.  } 
\end{figure*}

An interesting situation is presented in Fig. \ref{fig:Recoil}.  Due to the symmetry of the problem a reverse shock wave appears. Unlike the cartesian blast wave  case, here the two states separated by the contact discontinuity are not constant. It happens that in some localized regions the pressure is higher to the right side than to the left, producing a shock moving inwards. The velocities reached, for this reverse shock wave, are close to the speed of light. 

\begin{table}[htp]
\caption{\label{tab:BWinitialdata} Initial data configurations for the blast wave problem. Subindex $"i"$ is used to represent the initial primitive variables of the fluid in the inner chamber of radius $0.5$, whereas subindex $"e"$ is used to represent the variables in the outer chamber $r \in [0.5,1]$.}
\vspace{0.7cm}
\begin{tabular}{|l|l|l|l|l|l|l|}\hline 
Case		& $p_i$	&  $p_e$	& $\rho_i$  	& $\rho_e$  & $v_i$  & $v_e$ \\\hline
Weak blast 	& 1.0	&   0.1		& 1.0	    	& 0.125	& 0.0	& 0.0\\
Strong blast	& 133.33&   0.125	& 10.0		& 1.0	& 0.0	& 0.0\\\hline
\end{tabular}
\end{table}

\section{The evolution of a TOV star}
\label{sec:tov}

Concerning the evolution of the TOV star, besides the evolution of the fluid as described in detail in the preceding section, the space-time geometry is allowed to evolve according to Einstein's equations, therefore the coupled Euler-Eisntein system of equations holds.

Moreover, in this case the initial data one supplies is not arbitrary, instead they have to obey Einstein's field equations at initial time. 

In this section we construct various initial configurations and show how stable and unstable configurations behave.

\subsection{Einstein's equations}

Einstein's equations $G_{\mu\nu} = 8\pi T_{\mu\nu}$ for the space-time (\ref{eq:metric}) and the stress-energy tensor (\ref{eq:set_pf}) in terms of the conservative variables (\ref{eq:cons_vars}) reduce to the following system of equations, which coincide with those in \cite{HawkeMillmore} (see  \cite{MapleLeaf_eqs} for the details on the equations):

\begin{eqnarray}
\partial_t a &=& -4\pi r \alpha a S_r, \label{eq:Gtr}\\
\partial_r a &=& a^3 \left[ 4\pi r (\tau + D) - \frac{m}{r^2}\right] ,\label{eq:Gtt}\\
\frac{\partial_r \alpha}{\alpha} &=& a^2 \left[ 4\pi r (S_r v^r + p) + \frac{m}{r^2}\right] ,\label{eq:Grr}
\end{eqnarray}

\noindent where we have identified the metric function $a$ with the mass aspect function using the expression $a^2 = 1/(1-2m(r)/r)$, where $m(r)$ is the mass contained within a 2-sphere of radius $r$.

Notice that this is an overdetermined constrained evolution system, that is, the first equation is an evolution equation for the metric function $a$, the second is the Hamiltonian constraint and the third equation is a slicing condition for the lapse $\alpha$. This system of equations allows the evolution of any source provided $p$, $\rho$, $D$, $S_r$ and $\tau$, however in order to represent a solution of Einstein's equations they need to satisfy such equations at initial time, which we describe next.

\subsection{The initial value problem}

A TOV star is described as a spherically symmetric, static system that obeys Einstein's equations sourced by a perfect fluid that obeys a polytropic equation of state.

In order to solve the initial value problem we assume the space-time metric is static and momentarily will use $m(r)$ instead of $a$ in order to maintain the standard notation for the construction of TOV stars (see e.g. \cite{Baumgarte}). Then we start with the line element (\ref{eq:metric}) rewritten as:

\begin{equation}
ds^2 = -\alpha(r)^2 dt^2 + \frac{dr^2}{1-\frac{2m(r)}{r}} + r^2 d\theta^2 + r^2 \sin^2 \theta d\phi^2,
\end{equation}

\noindent where we have assumed the system is time-symmetric at $t=0$ and the metric functions and the gas functions depend only on $r$. We also assume the gas obeys initially a polytropic equation of state $p=K\rho ^{\Gamma}$. Using \cite{MapleLeaf_ivp} one arrives at the following set of equations:

\begin{eqnarray}
\frac{dm}{dr} &=& 4\pi r^2 \rho, \label{eq:tov_mass}\\
\frac{dp}{dr} &=& -\frac{m}{r^2}(\rho+p)\left( 1 + \frac{4\pi r^3 p}{m} \right) \left( 1 - \frac{2m}{r} \right)^{-1}\nonumber\\
&=& -(\rho + p) \frac{m+4\pi r^3 p}{r(r-2m)},\label{eq:tov_p}\\
\frac{1}{\alpha}\frac{d\alpha}{dr} &=& -\frac{1}{\rho+p} \frac{dp}{dr}= \frac{m+4\pi r^3 p}{r(r-2m)},\label{eq:tov_alpha}
\end{eqnarray}

\noindent where we have used $\rho=\rho_0 (1+\epsilon)$ and $\rho_0 h = \rho_0(1+\epsilon)+p=\rho+p$. This system of ordinary equations constitutes the conditions a TOV star satisfies at initial time, and has to be integrated outwards from $r=0$ up to $r=r_{max}$. We solve these equations using a fourth order Runge-Kutta integrator on top of the same grid defined for the fluid equations described above.
The initial conditions for the integration of the variables are:

\begin{enumerate}
\item $m(0)=0$, because the integrated mass up to there is zero. Another interpretation is that the gravitational field at the origin is zero, and thus $a(r)=1$ corresponds to the flat space, which implies $m(0)=0$.

\item $p(0)=K\rho_{c}^{\Gamma}$, where $\rho_c$ is the central value of the rest mass density. In the whole domain it happens that on the one hand $\rho_0 = (p/K)^{1/\Gamma}$ and on the other, from an ideal gas equation of state $p = (\Gamma-1)\rho_0 \epsilon, ~\Rightarrow ~ \rho_0 \epsilon = p/(\Gamma-1)$; therefore $\rho = \rho_0 (1+\epsilon) = \rho_0 + \rho_0\epsilon = (p/K)^{1/\Gamma} + p/(\Gamma -1)$ is the source of (\ref{eq:tov_mass}).

\item $\alpha(0) = \alpha_0$ is an arbitrary given initial central value for the lapse. Notice in (\ref{eq:tov_alpha}) that the solution can be rescaled multiplying by a constant, which preferably will be chosen such that at the numerical boundary satisfies $\alpha(r_{max})=1/a(r_{max})$, which is a condition that Schwarzschild's solution satisfies and we expect to happen at $r=r_{max}$.

\item The value of $\rho_c$ turns to be the input parameter that determines the configuration, and corresponds to the central value of the total energy density. The result is that for each value of $\rho_c$ a configuration can be constructed.
\end{enumerate}

Two observations are in turn. The first one concerns the point $r=0$, because there equations (\ref{eq:tov_p}) and (\ref{eq:tov_alpha}) are singular. What is usually done is to Taylor expand the singular factor and get approximate equations for small values of $r$:

{\small
\begin{eqnarray}
&&\frac{m+4\pi r^3 p}{r^2 - 2mr} \nonumber\\
&\sim& \frac{m(0)+m'(0)r + \frac{1}{2}m''(0)r^2 + \frac{1}{6}m'''(0)r^3+O(r^4) + 4\pi r^3 p}{r^2 - 2r(m(0) +m'(0)r + \frac{1}{2}m''(0)r^2 + \frac{1}{6}m'''(0)r^3 + O(r^4))} \nonumber\\
&=& \frac{4\pi \rho r/3 + 4\pi r p}{1 - 8\pi \rho r^2 / 3}
\end{eqnarray} 
}

\noindent where equation (\ref{eq:tov_mass}) was used to calculate the derivatives of $m$: $dm/dr |_{r=0} = 4\pi r^2 \rho |_{r=0} = 0$, $d^2 m/dr^2 |_{r=0} = 4\pi(2r\rho + r^2 \rho)|_{r=0} = 0$ and $d^3 m/dr^3|_{r=0} = 4\pi (2\rho + 2r\rho + r^2 \rho)|_{r=0}=8\pi\rho$. Then equations (\ref{eq:tov_p}) and (\ref{eq:tov_alpha}) are approximated for small $r$ by

\begin{eqnarray}
\frac{dp}{dr} &=& - (\rho+p) \frac{4\pi \rho r /3+ 4\pi r p}{1 - 8\pi \rho r^2/3},\nonumber\\
\frac{1}{\alpha}\frac{d\alpha}{dr} &=&  \frac{4\pi \rho r /3+ 4\pi r p}{1 - 8\pi \rho r^2/3}.\label{eq:regularized}
\end{eqnarray}

\noindent These approximate regular equations are the ones to be programmed for at least the first mesh point located at $r=\Delta r$ where $\Delta r$ is the spatial resolution of the mesh.

The second observation is related to the divergence of the specific enthalpy (\ref{eq:enthalpy}) when $\rho_0$ approaches zero, which in theory would happen from the star's surface to infinity where there is only vacuum. It is usually set an external atmosphere, that is, a minimum value is assumed for $\rho_0$ than can be hidden within numerical errors and allows the convergence of the numerical calculations, however it happens to be a mere numerical artifact at the moment and as far as we can tell,  there is no theory behind the appropriate value of the atmosphere density $\rho_{atm}=floor$. The value of $floor$ rather depends on the specific problem to be solved.

Considering this ingredient one can define the radius $R$ of the TOV star as the minimum radius $r=R$ at which $\rho_0 = floor$. On the other hand, the total mass of the TOV star is $M_T=m(R)$, whereas the rest mass of the star is the spatial integral of $\rho_0$ given by $M_0=4\pi \int^{R}_{0}\rho_0 r^2 a(r) dr$. The difference between the total and the rest mass of the star determines whether or not the system is gravitationally bounded.

\begin{figure}[htp]
\includegraphics[width=8cm]{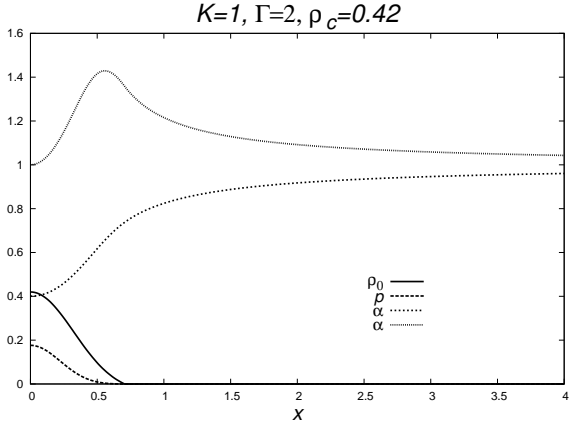}
\caption{\label{fig:scaled_id} Initial data as found for $\Gamma=2, ~(n=1)$, $\rho_c = 0.42$, $\alpha_c=0.5$, $K=1$. The result is as follows, the total mass $M=0.1616$, the rest mass $M_0=0.177$, $R=0.7045$. The lapse $\alpha$ has been rescaled such that $\alpha(r_{max})=1/a(r_{max})$ as expected to happen for Schwarzschild's solution.}
\end{figure}

As an example of a TOV star configuration we show in Fig. \ref{fig:scaled_id} the functions for $\Gamma=2$, a polytropic constant $K=1$ and a central density $\rho_c=0.42$.

\begin{figure}[htp]
\includegraphics[width=8cm]{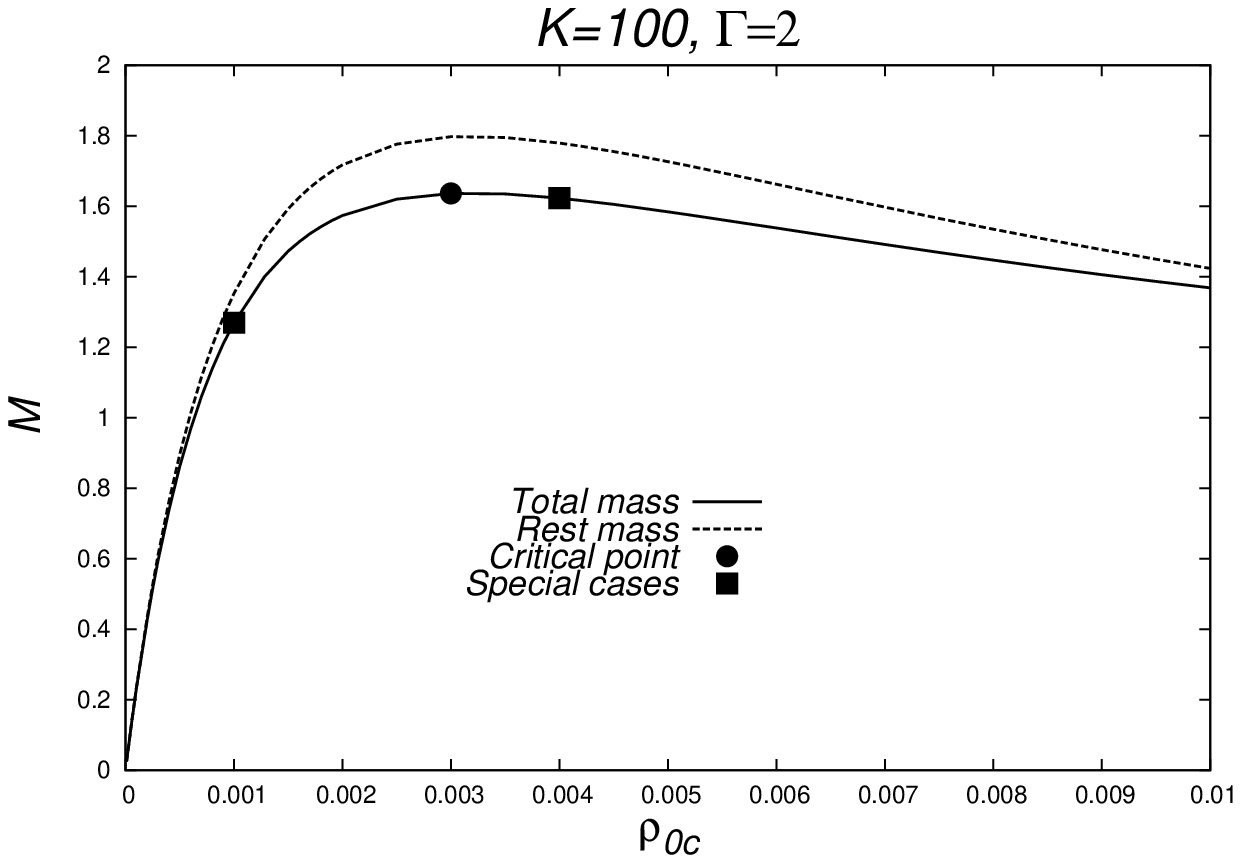}
\includegraphics[width=8cm]{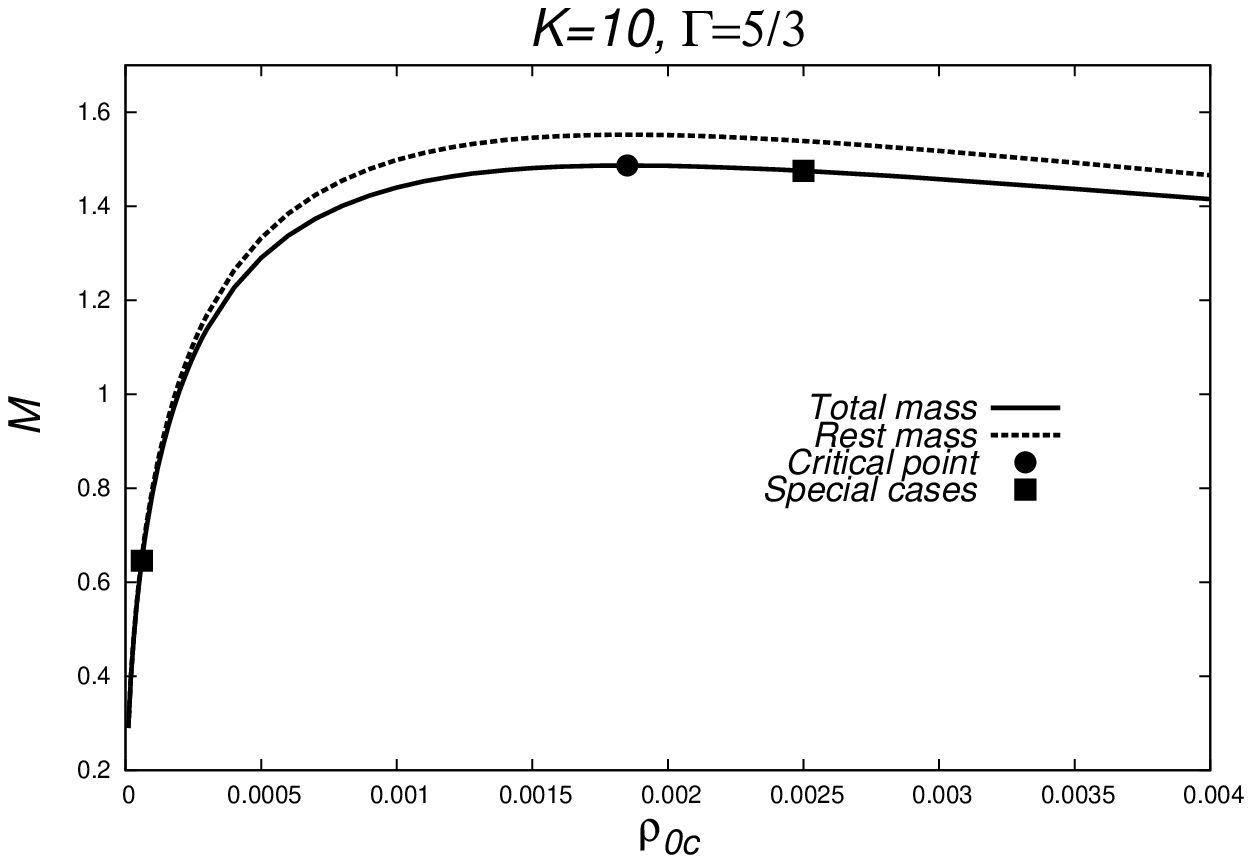}
\caption{\label{fig:mass_vs_rho} Mass vs central density diagram for the cases $K=100$, $\Gamma=2$ and $K=10,\Gamma=5/3$. The filled circle indicates the location of the maximum possible mass and also the threshold between the stable and unstable brances. Those configurations to the left of the maximum are stable and those to the right are unstable. The fact that the rest mass has a bigger value than the total mass indicates that the system is gravitationally bounded. This eventually implies that those configurations belonging to the unstable branch should collapse and form black holes. Configurations marked with a filled square correspond to particular configurations we evolve to illustrate the different behaviors of stable and unstable configurations.}
\end{figure}

The result of integrating the TOV equations for various values of $\rho_c$ is summarized in Fig. \ref{fig:mass_vs_rho}, where we plot the total and rest mass for two different classes of equations of state, $\Gamma=2$ and $\Gamma=5/3$ for several values of $\rho_c$. Each point in the curves corresponds to a value of $\rho_c$ and therefore defines a TOV star configuration. The first plot corresponds to an ultrarelativistic case whereas the second serves to model a fermionic gas and is a simple approximate model of white dwarfs. The maximum in the plots indicates the threshold between stable and unstable configurations, that is, configurations to the left of the maximum oscillate under perturbations whereas those to the right collapse and form black holes if they are perturbed, because these systems are gravitationally bound since $M_0>M_T$ for the values of $\rho_c$ shown. We also indicate in Fig. \ref{fig:mass_vs_rho} four particular configurations, two stable and two unstable that we use to illustrate their evolution.

Summarizing, the information required to start up the evolution of a TOV star has now been calculated at each cell, and is the following:

\begin{eqnarray}
a &&~~~ (numerically~ integrated),\nonumber\\
\alpha &&~~~ (numerically~integrated),\nonumber\\
p && ~~~(numerically~integrated),\nonumber\\
\rho_0 &=& max(\left(\frac{p}{K}\right)^{1/\Gamma},floor),\nonumber\\
v^r &=& 0,\nonumber\\
W &=& \frac{1}{\sqrt{1-a^2 (v^r)^2}} ,\nonumber\\
\epsilon &=& \frac{p}{\rho_0 (\Gamma -1)},\nonumber\\
h &=& 1 + \epsilon + p/\rho_0,\nonumber\\
D &=& \rho_0 W, \nonumber\\
S_r &=& \rho_0 h W^2 a^2 v^r, \nonumber\\
\tau &=& \rho_0 h W^2 -p -\rho_0 W.\nonumber
\end{eqnarray}

\noindent Then the evolution of the system is ruled by the Einstein-Euler system as described next.

\subsection{The evolution}

The system of equations is the one composed of Euler's equations (\ref{eq:FluxConservative2}) and the overdetermined system of Einstein's equations (\ref{eq:Gtr}-\ref{eq:Grr}). The whole system is started  with the initial data corresponding to a TOV star. Among Einstein's equations we choose to solve (\ref{eq:Gtr}) for $a$ and the remaining (\ref{eq:Gtt}) is the Hamiltonian constraint we use to monitor the evolution. Notice that the equation for $\alpha$ is an ODE in $r$, that we integrate every time step during the evolution.

\begin{figure*}[htp]
\includegraphics[width=8cm]{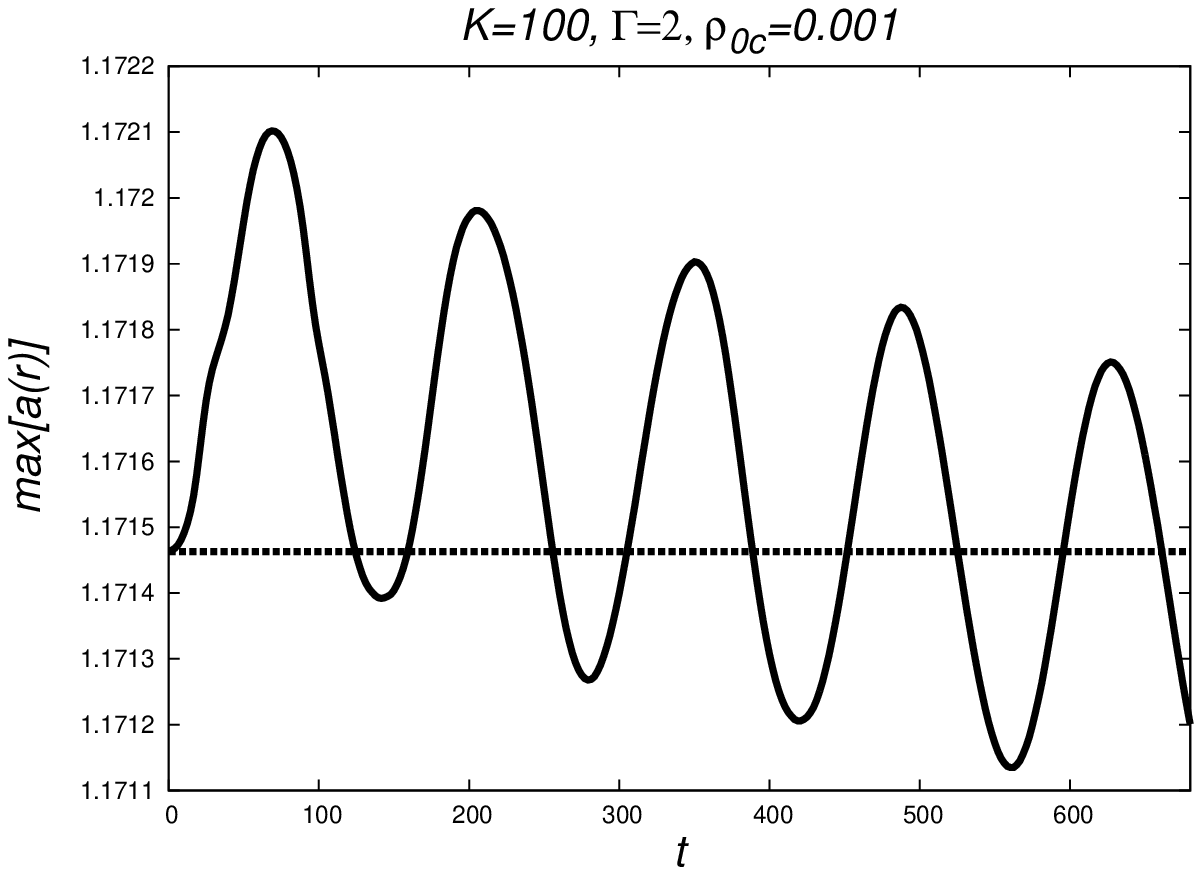}
\includegraphics[width=8cm]{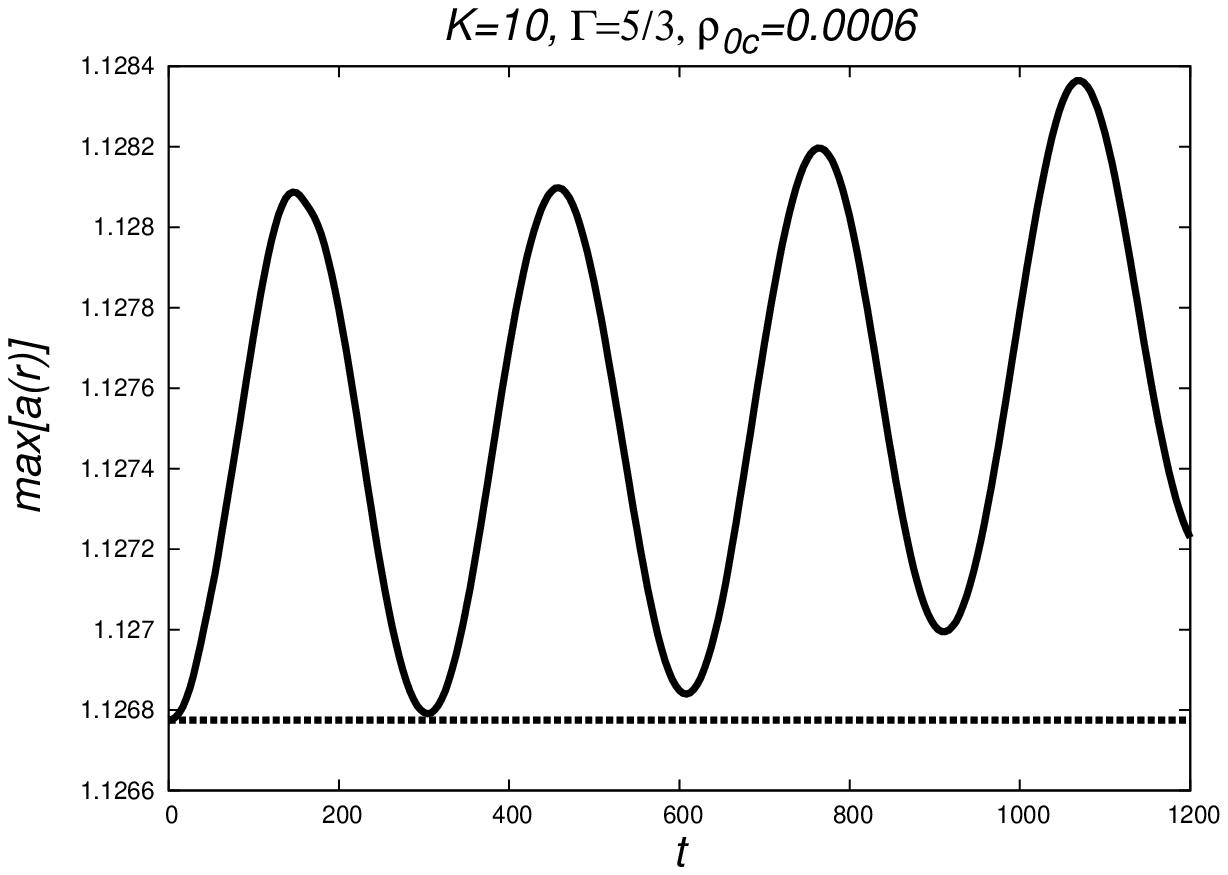}
\includegraphics[width=8cm]{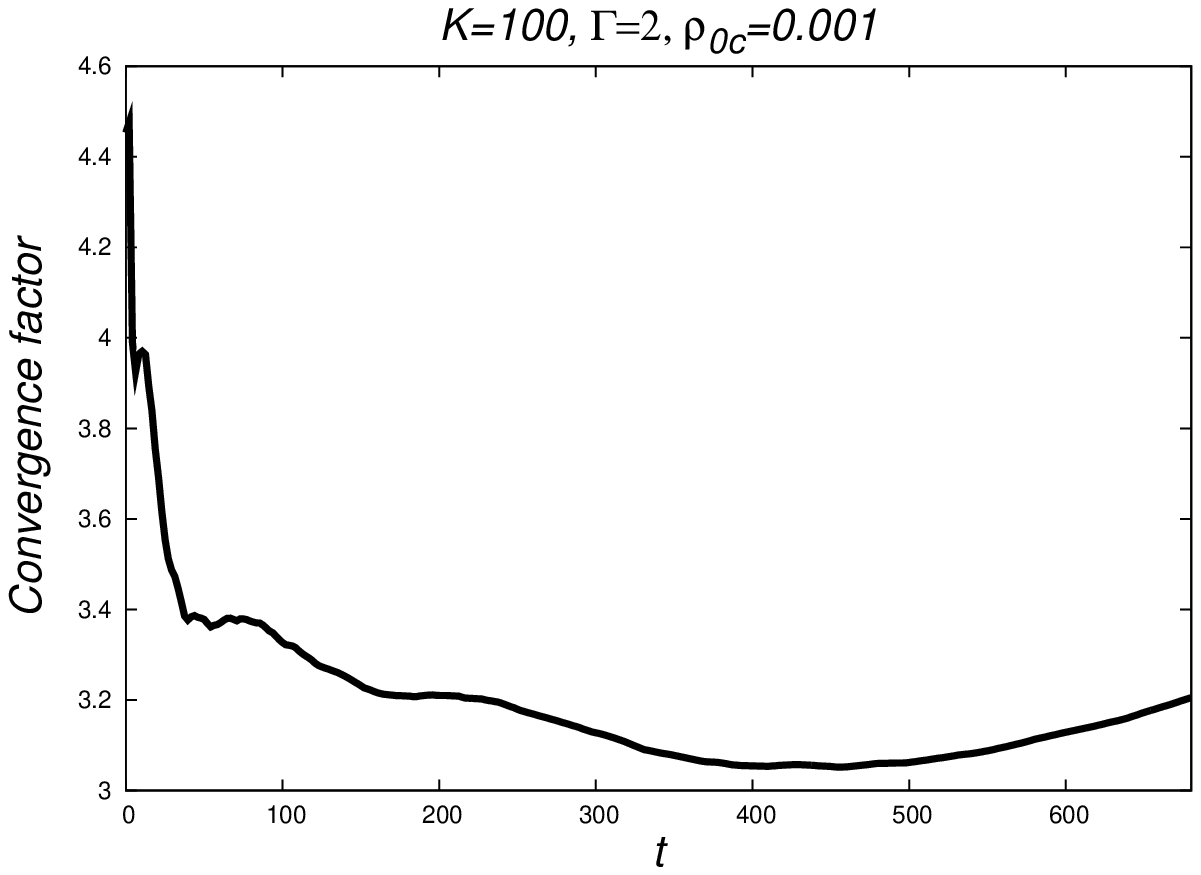}
\includegraphics[width=8cm]{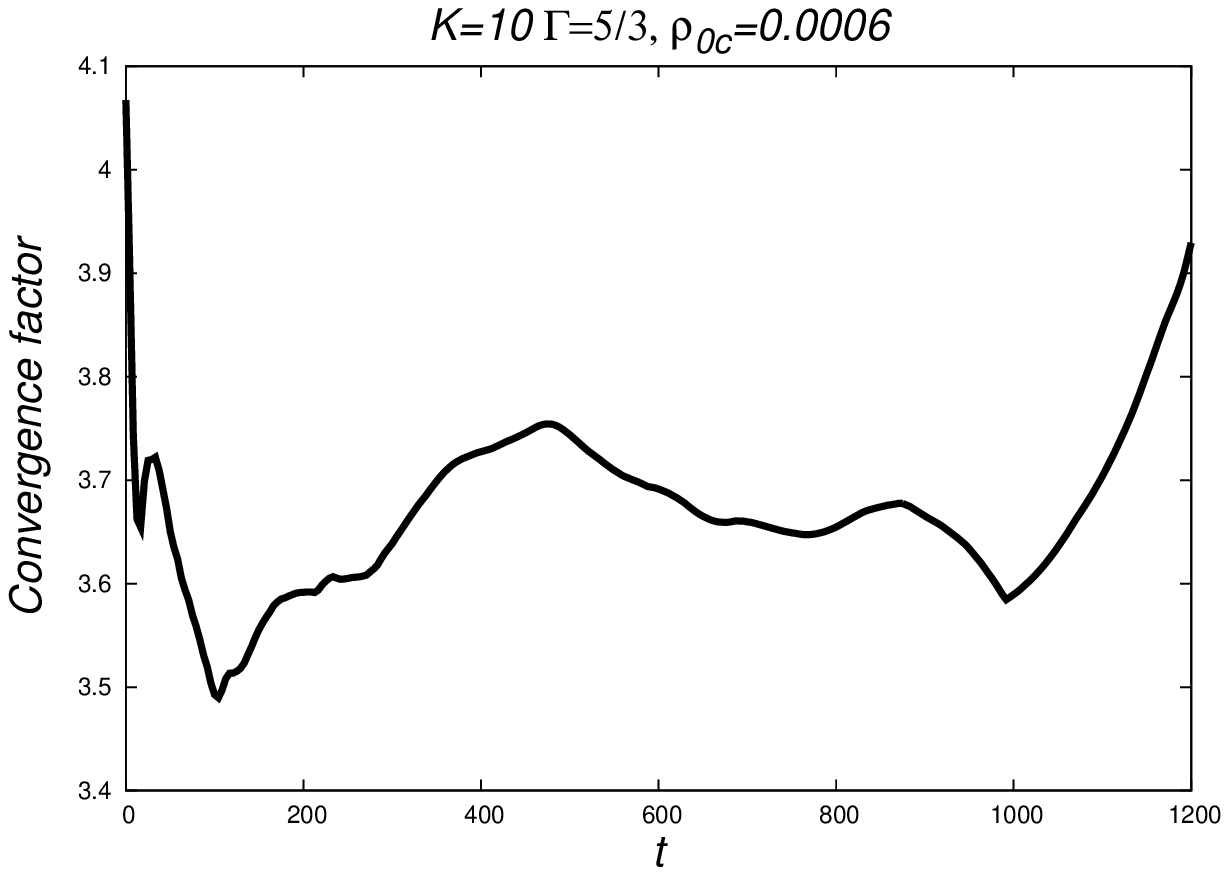}
\caption{\label{fig:stable} Maximum of $a(r)$ for the stable cases $K=100$, $\Gamma=2$ and $\rho_c=0.001$ and $K=10$, $\Gamma=5/3$ and $\rho_c=0.0006$. The metric function responds to the perturbation due to numerical errors and its maximum oscillates.  The constant line indicates the value calculated at initial time for the maximum of $a$ which should be maintained constant in the continuum limit. Also shown is the convergence factor  of the $L_1$ norm of the Hamiltonian constraint of Einstein's equations. This simulations use 6000 cells in a domain $r \in [0,500]$ for the $\Gamma=2$case and 3000 cells for the case $\Gamma=5/3$ in a domain $r \in [0,500]$. In both cases the atmosphere rest mass density is $floor=10^{-13}$.}
\end{figure*}

\begin{figure*}[htp]
\includegraphics[width=8cm]{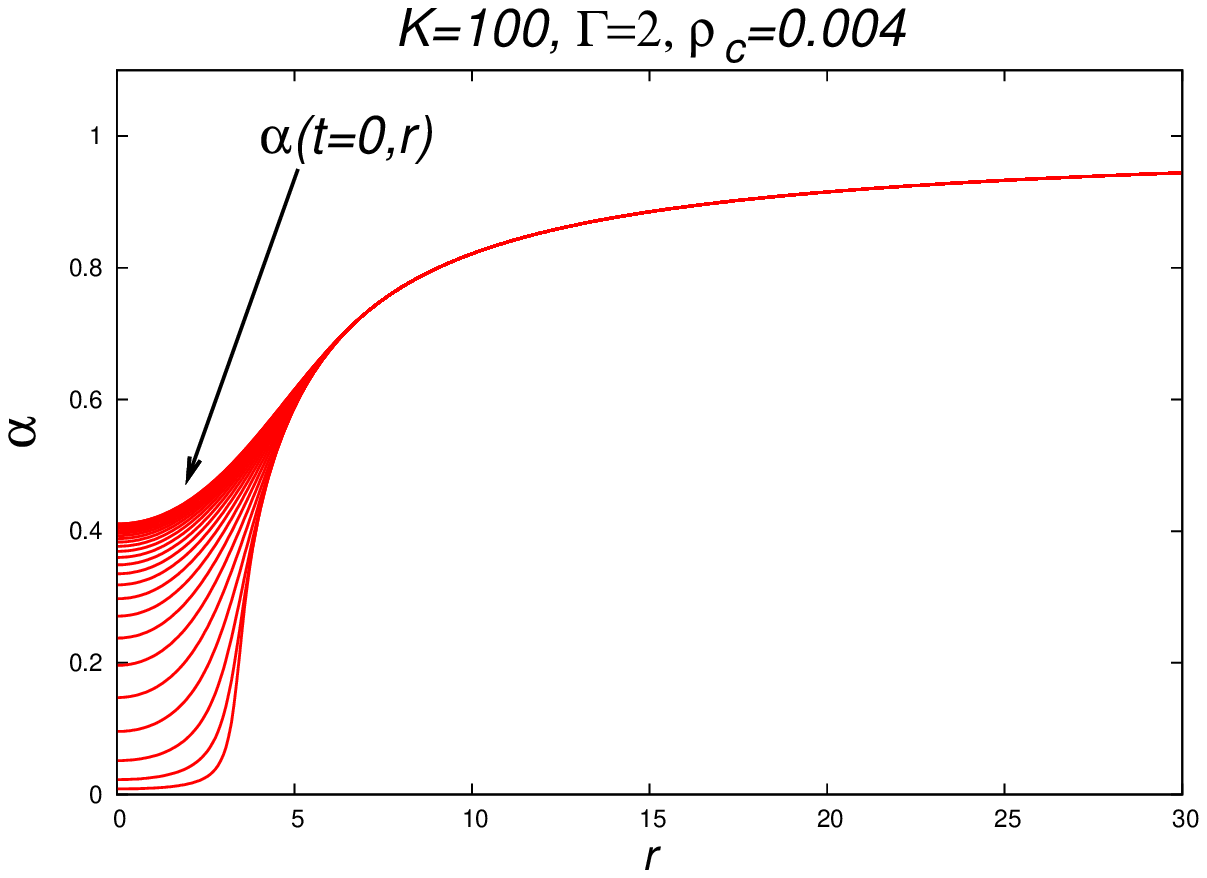}
\includegraphics[width=8cm]{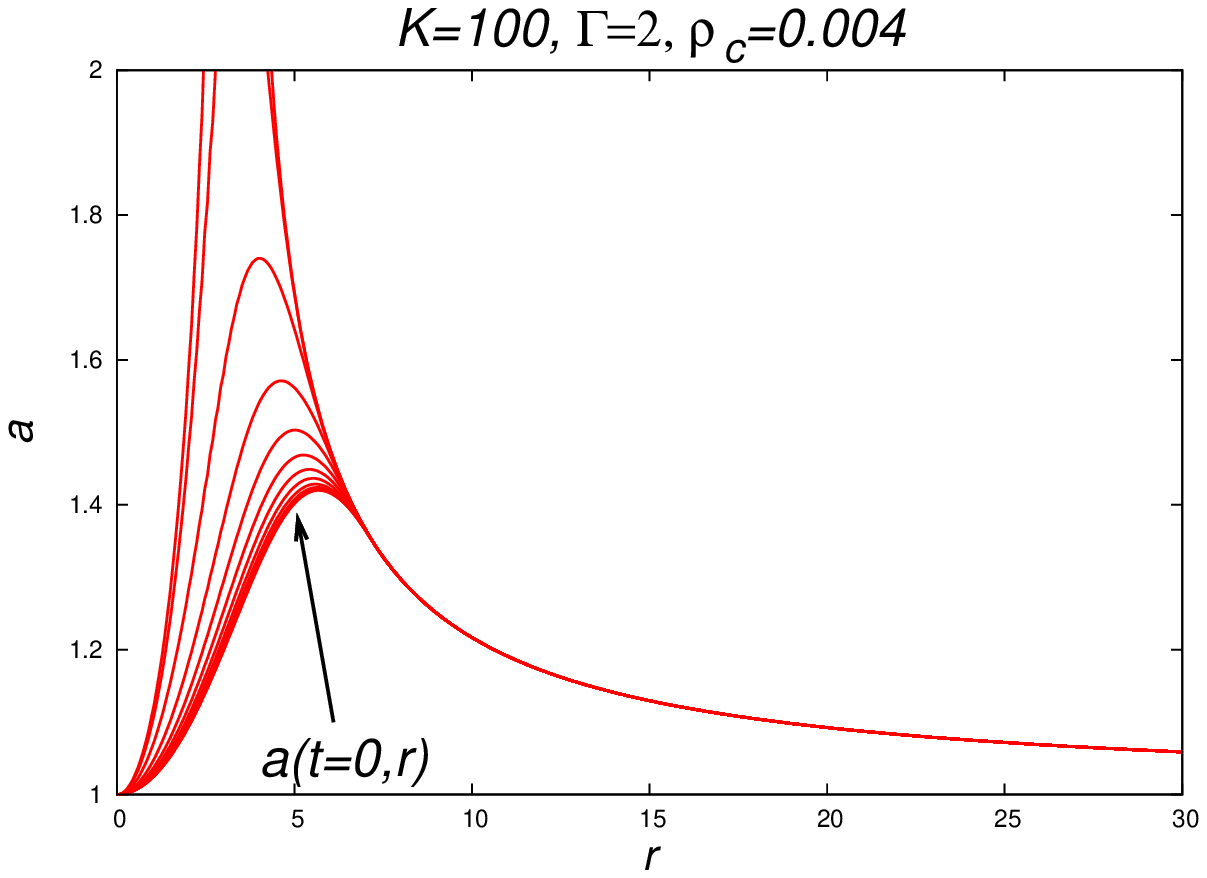}
\includegraphics[width=8cm]{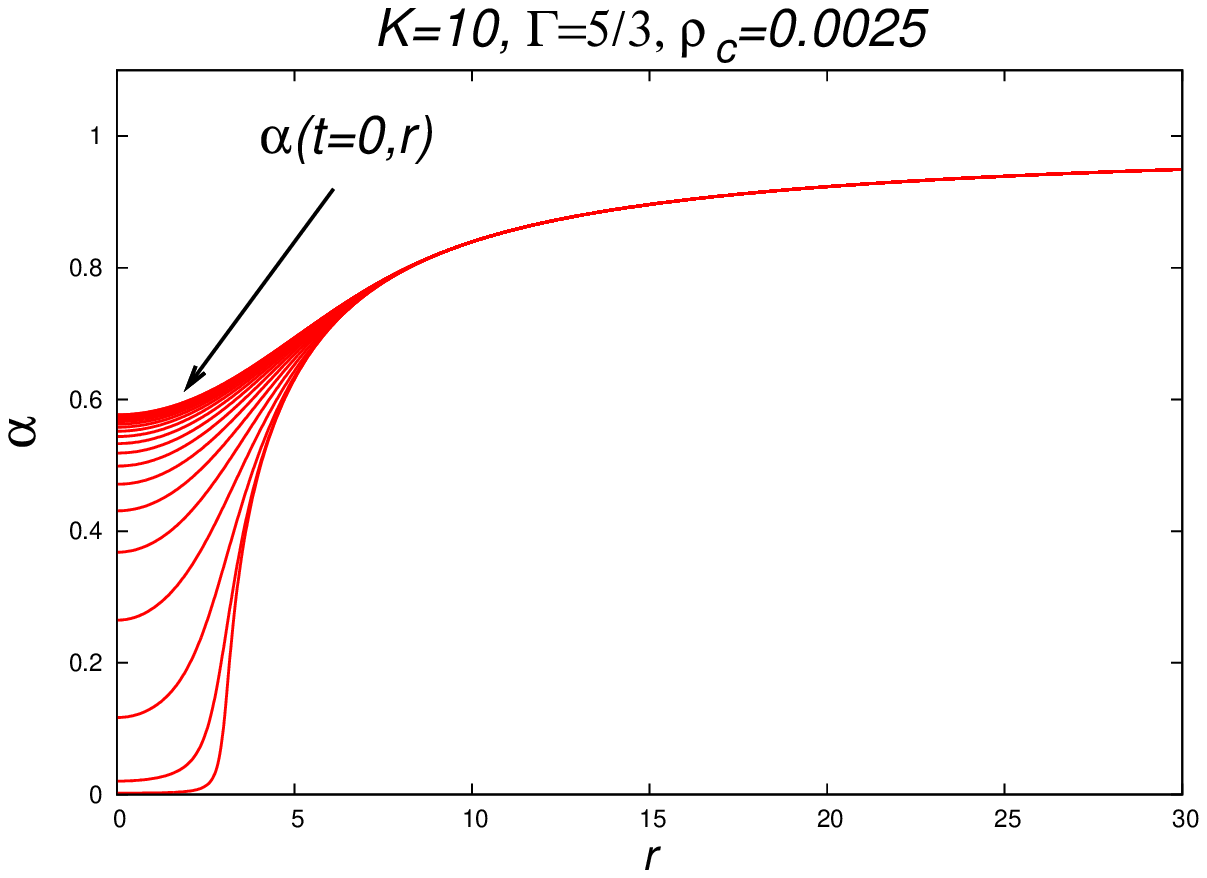}
\includegraphics[width=8cm]{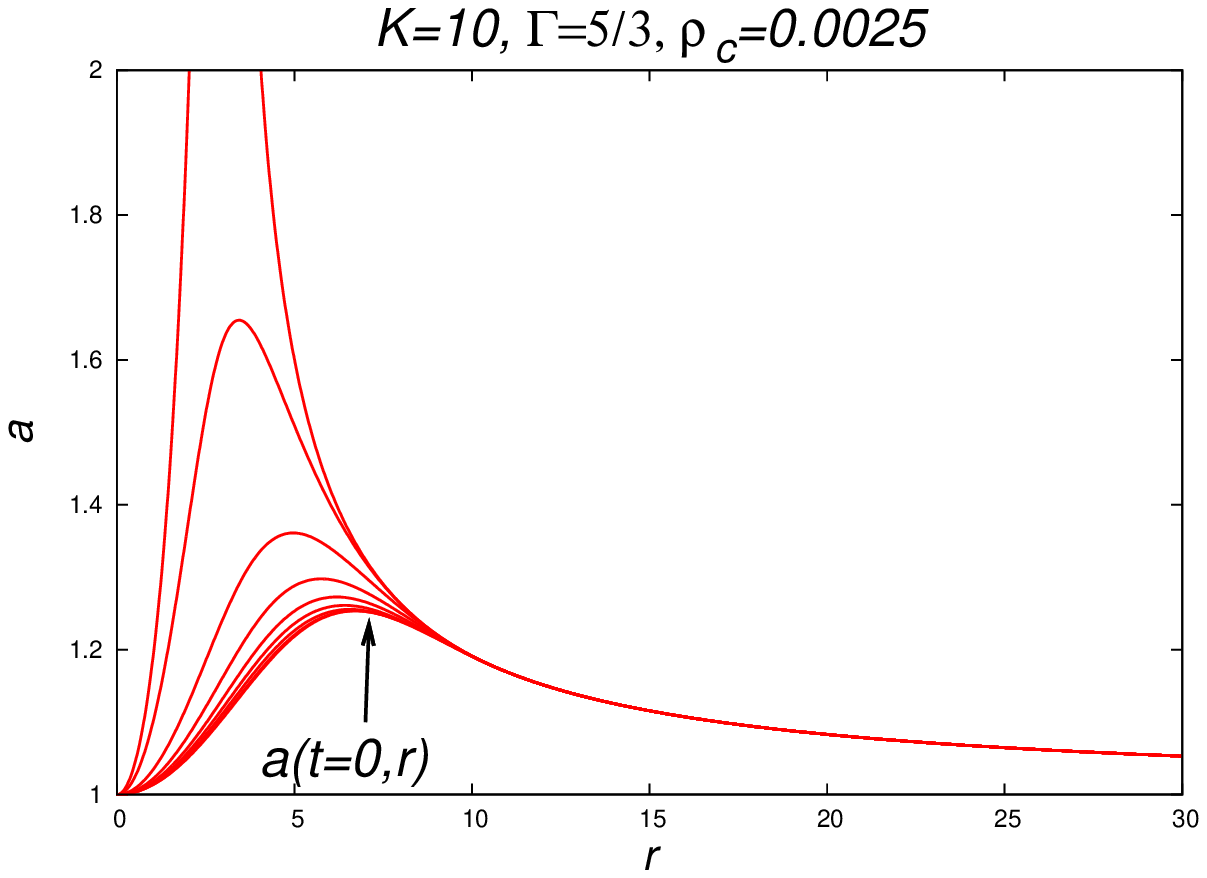}
\caption{\label{fig:unstable} We show snapshots of the metric functions for the unstable cases indicated in Fig. \ref{fig:mass_vs_rho}, $(K=100, ~\Gamma=2,~\rho_c=0.004)$ with total mass $M_T=1.623$ and $(K=10, ~\Gamma=5/3,~\rho_c=0.0025)$ with $M_T=1.475$. The lapse collapses to zero with time, which indicates that an apparent horizon has formed, and in turn implies that external to such apparent horizon there is an event horizon; observe that the lapse approaches zero until $r\sim 2M_T$. Notice also that the metric function $a$ diverges at a similar location of the horizon radius, which is an effect of the slice stretching that occurs during a black hole formation in non-penetrating coordinates.}
\end{figure*}

\begin{figure}[htp]
\includegraphics[width=8cm]{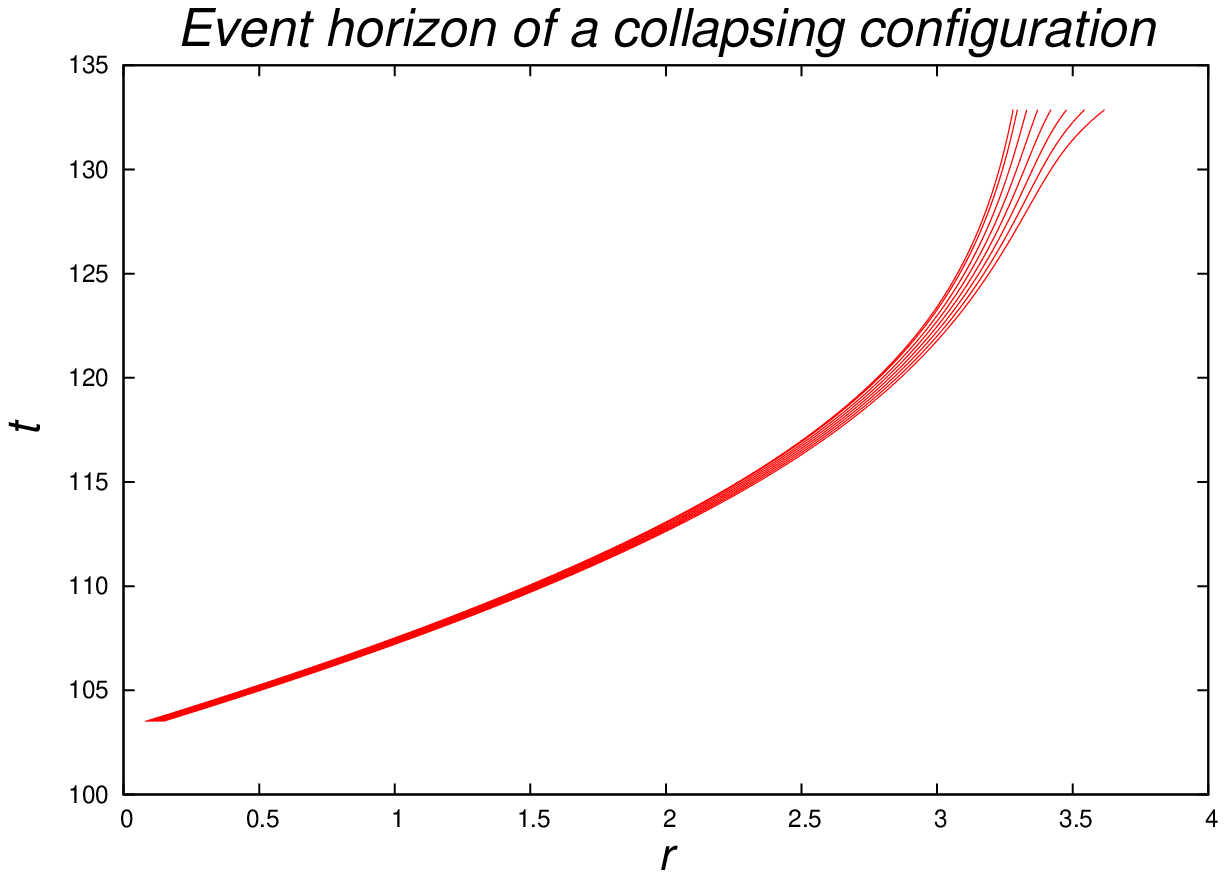}
\caption{\label{fig:eh} We show a bundle of null rays, which due to the symmetry and coordinates we are using, represent the behavior of 2-dimensional null spheres. What we show is that some of these rays diverge toward the singularity and other will diverge outwards toward infinity, and more precisely toward future null infinity \cite{Hawking}. The growing null sphere in between the two behaviors would be the event horizon. In theory, an event horizon would be well defined if we guarantee that outgoing null rays diverging outwards reach future null infinity, however we only show a small chunk of the space-time where we can measure the divergence of these null surfaces due to the aforementioned slice stretching drawback of the coordinates we are using.}
\end{figure}

Thus the algorithm for the evolution provided given values of $K$ and $\Gamma$ is as follows:

\begin{itemize}
\item Construct a TOV configuration as initial data for a given central density $\rho_c$.
\item Use the evolution equations and calculate new $D$, $S_r$, $\tau$ and simultaneously integrate in time the momentum constraint (\ref{eq:Gtr}) for $a$ at every time step.
	\begin{itemize}
	\item[] For each intermediate time-step of the MoL integrator
	\item Express the primitive variables in terms of the conservative variables and  reconstruct  to the 			left and right from intercell boundaries the values for $p$, $v^r$, $	\rho_0$,  and the conservative 			variables $D,~S_r,~\tau$ in order to obtain the necessary information to construct the 				numerical fluxes.
	\item Apply boundary outflux conditions to the conservative variables and extrapolate for $a$. In our conservative formulation it requires only to copy the values of the conservative variables at  boundaries from the point next to it.
	\item Integrate (\ref{eq:Grr}) for $\alpha$, and rescale it such that at the boundary it 			satisfies $\alpha(r_{max}) = 1/a(r_{max})$.
	\end{itemize}
	\item After a full time step, calculate the Hamiltonian constraint (\ref{eq:Gtt}) in order to monitor the 			convergence of the results.
\end{itemize}

Let us explain what should happen during the evolution. In the continuum limit the TOV configurations should remain time independent all the way, because they are static solutions to Einstein's equations. If a perturbation is applied (for example, a small amplitude shell pulse added to the density), the geometry and matter quantities would oscillate around the equilibrium values, whereas an unstable configuration would collapse and form a black hole. Nevertheless, we are using numerical methods and as shown above, all our calculations involve an intrinsic error. We then take advantage of such fact and use such error as the perturbation of the equilibrium configurations. Therefore stable configurations would oscillate around the equilibrium values, whereas unstable configurations eventually will collapse due to a perturbation triggered by the numerical errors.

In order to illustrate the evolution of TOV stars we choose two stable configurations and show some results in Fig. \ref{fig:stable}. On the one hand we show the maximum of the metric function $a$ in time which shows a periodic oscillation. The reason is that we are solving numerically the initial value problem, and also we are integrating numerically with a finite accuracy, then there is a numerical error introduced in our calculations at initial time which works as a perturbation whose effects converge to zero in the continuum limit \cite{Guzman2010}. What is more important is that the metric function remains nearly time-independent, as expected for a stable equilibrium configuration. On the other hand we show the convergence of the Hamiltonian constraint, which is necessary to verify that we are truly solving the full set of Einstein's equations (remember that the Hamiltonian constraint (\ref{eq:Gtt}) is not being solved, only monitored). We are verifying the convergence of our results by doubling the resolution, which means that a convergence factor is defined as $2^Q$ where $Q$ is the order of convergence \cite{Guzman2010}. Then from Fig. \ref{fig:stable} we know our results converge within order 1.6 and order 2, which is consistent with the approximations we have made in all the methods used.

We also show the evolution of two unstable configurations in Fig. \ref{fig:unstable}. In this case the metric functions $\alpha$ and $a$ do not remain nearly time independent as in the stable cases, where snapshots of the metric functions would be seen as a single curve. Instead, the lapse collapses to zero in a localized region, which in the coordinates we use means that an apparent horizon and that a black hole has been formed. Also the function $a$ diverges near the location of the horizon, which is due to the slice stretching effect of the normal coordinates we are using. 

In order to make sure that a black hole has formed we track a bundle of outgoing null geodesics starting at about $t\sim103$, which we show in Fig. \ref{fig:eh} for one of the collapsing configurations. The null rays shown indicate the behavior of null 2-spheres. Near the event horizon these null surfaces should diverge toward the singularity and toward future null infinity. We show the null geodesics until our simulation remains accurate, which happens until the aforementioned problem of the coordinates we are using appears. However this small window in time allows one to appreciate the divergence of the null spheres and thus infer that the event horizon is contained into the set of null rays shown. This guarantees that not only an apparent horizon has been formed through the gauge dependent condition $\alpha \sim 0$ but also an event horizon, which is gauge independent. It is possible to see that the event horizon grows due to the accretion of the gas and tends to stabilize at a radius nearly twice the mass of the initial configuration.


\section{Final comments}
\label{sec:comments}

We have shown in detail a particular sort of implementation of numerical relativistic hydrodynamics solutions, of spherically symmetric cases in spherical coordinates. The steps specified in the paper are also useful for different choices of numerical approximations described here. 

Specifically, related to the treatment of relativistic hydrodynamics, we only use a particular flux formula for the numerical solution of the Riemann problems at the intercell boundaries. There are several other choices like the Roe, Marquina, HLL, HLLC,etcetera flux formulas. Also, for the cell reconstruction of variables, other choices aside the minmod limiter are well studied like the MC (linear monotonic centered), PPM (parabolic piecewise method), etc.

For instance in \cite{HawkeMillmore} the authors use the combination HLLE flux formula and MC limiter for the evolution of TOV stars, and in \cite{Font2001} the evolution of TOV stars is implemented using Marquina and Roe fluxes with MC and minmod slope reconstructors.
 
We found appropriate to choose a single combination of numerical methods in order to be as specific and detailed as possible.

We also want to mention other aspects inherent to these numerical methods. Particularly interesting is that the enthalpy diverges when the rest mass density approaches zero (\ref{eq:enthalpy}), and the implementation of an atmosphere is necessary. However, so far there is no theory or explanation about what values of the atmosphere density are to be applied, and in the best cases (as here) convergence tests are used to support the numerical results, and the values used for such external density is justified as long as the numerical results in terms of accuracy and convergence are achieved. A potential recipe for an atmosphere with a different equation of state may ameliorate this problem \cite{HawkeMillmore}.

Even though the density at the atmosphere is small, the fluid may develop highly relativistic speeds, which eventually may produce intractable shocks at the star surface. Therefore the atmosphere requires a rather ad hoc treatment, like artificial limitation to the speed of the fluid at the atmosphere, etc. In particular in our simulations of TOV stars we have used this type of condition.


\section*{Acknowledgments}

This work is supported by grants
CIC-UMSNH-4.9 
and CONACyT 106466. FDLC and MDMA acknowledge support from CONACyT.



\end{document}